\documentclass{amsart}
\usepackage{amssymb,stmaryrd}
\usepackage{amsfonts}
\usepackage{amstext}
\usepackage{algorithmic}
\usepackage{algorithm}
\usepackage{graphicx}
\usepackage{epstopdf}
\usepackage[all]{xy}
\usepackage{MnSymbol}
\usepackage{tikz}
\usepackage{braket}

\numberwithin{equation}{section}
\newtheorem{theorem}{Theorem}[section]

\newtheorem{proposition}[theorem]{Proposition}

\theoremstyle{definition}

\theoremstyle{remark}

\newcommand{\R}{{\mathbb{R}}}

\newcommand{\C}{{\mathbb{C}}}
\newcommand{\Z}{{\mathbb{Z}}}

\newcommand{\N}{{\mathbb{N}}}

\newcommand{\wedgeq}{{\wedge\kern-5pt\cdot}}

\newcommand{\tens}{\otimes}

\newcommand{\id}{{\rm id}}

\newcommand{\extd}{{\rm d}}
\newcommand{\del}{{\partial}}
\newcommand{\eps}{\epsilon}

\begin{document}

\title{Fuzzy and discrete black hole models}
	
\thanks{Ver1 The first author was partially supported by CONACyT (M\'exico)}
	
\author{J. N. Argota-Quiroz and S. Majid}
\address{Queen Mary, University of London\\
		School of Mathematics, Mile End Rd, London E1 4NS, UK}
	
\email{j.n.argotaquiroz@qmul.ac.uk, s.majid@qmul.ac.uk}
	
\begin{abstract} Using quantum Riemannian geometry, we solve for a ${\rm Ricci}=0$ static spherically-symmetric solution in 4D, with the $S^2$ at each $t,r$ a noncommutative fuzzy sphere,  finding a dimension jump with solutions having the time and radial form of a classical 5D Tangherlini black hole. Thus, even a small amount of angular noncommutativity leads to radically different radial behaviour, modifying the Laplacian and the weak gravity limit. We likewise provide a version of a 3D black hole with the $S^1$ at each $t,r$ now a discrete circle $\Z_n$, with the time and radial form  of the inside of a classical 4D Schwarzschild black hole far from the horizon. We  study the Laplacian and the classical limit $\Z_n\to S^1$. We also study the 3D FLRW model on  $\R\times S^2$ with $S^2$  an expanding fuzzy sphere and find that the Friedmann equation for the expansion is the classical 4D one for a closed $\R\times S^3$ universe.  \end{abstract}

\maketitle 
	
\section{Introduction}

The idea that not only quantum phase spaces but spacetime coordinates themselves could be noncommutative or `quantum' due to quantum gravity effects has been around since the first days of quantum theory. An often cited early work was \cite{Sny}, although not proposing a closed spacetime algebra as such. In modern times, such a {\em quantum spacetime hypothesis} was proposed in \cite{Ma:pla} on the grounds that the division into position and momentum should be arbitrary and hence if these do not commute then so should position and momentum separately noncommute. Several flat quantum spacetimes were studied in the 1990s\cite{DFR,MaRue,Hoo}, but only recently has there emerged a constructive formalism of quantum Riemannian geometry\cite{BegMa} to make possible curved models\cite{BegMa:gra,MaTao,Ma:sq,Ma:haw,ArgMa,LirMa}. This formalism is rather different in from Connes' well-known `spectral triple' approach to noncommutative geometry\cite{Con} based on an axiomatically defined `Dirac operator', coming more out of experience with quantum groups (but not limited to them).

In the present work, we take curved quantum spacetime model building to the next level with   black hole and FLRW cosmological models. Here, \cite{ArgMa}, introduced an expanding FLRW model based on $\R\times S^1$ with $S^1$ replaced by a discrete group $\Z_n$ with noncommutative differentials, while the coordinate algebra itself remains commutative. In Section~\ref{secFLRW}, we similarly look at the 3D $\R\times S^2$ model but replace $S^2$ by a noncommutative fuzzy sphere, defined as the usual angular momentum algebra at a fixed value of the quadratic Casimir (a coadjoint orbit quantisation) and with differential structure as recently introduced in \cite{BegMa}. We then proceed to our main results,  noncommutative black hole models.  Previously,  4D black holes were studied in a semidirect `almost commutative' quantisation\cite{Ma:alm} but with the quantum geometry only implicitly defined through the wave operator constructed as a noncommutative extension to the classical differential calculus. Also previously, a 4D FLRW model was constructed in a deformation setting at the Poisson-Riemannian geometry level\cite{FriMa} and with the  expected dimension, but at the price of nonassociativity of the exterior algebra due to curvature of the Poisson connection. Hence the models in the present paper are the first fully noncommutative FLRW and black hole ones that we are aware of within usual (associative) quantum Riemannian geometry.  Section~\ref{secBHfuz} does the 4D black hole with $S^2$ in polars replaced by a fuzzy one, while Section~\ref{secBHZn} looks for a 3D black hole model with the $S^1$ in polar coordinates replaced by $\Z_n$. The latter is not flat but Ricci flat (which can not happen classically in 3D) and has a naked singularity rather than a horizon.

A common feature that we find is what we call `dimension jump'. It has long been known that quantum exterior algebras often have extra dimensions that could not be predicted classically. Thus, in \cite{Ma:haw,ArgMa} the calculus on $\Z$ and $\Z_n$ respectively was actually 2D and this made possible curvature effects not expected for a classical line or circle. The limit  of the geometry on $\Z_n$ as $n\to \infty$ was likewise shown in \cite{ArgMa}  to be a classical circle but with an extra 1-form $\theta'$ normal to the circle when viewed in the plane. We then found that the Friedmann equations for the discrete $\R\times\Z_n$ model were the same as for a classical flat $\R\times \R^2$ FLRW model (an expanding plane), i.e. a dimension jump. The same will apply now for the fuzzy sphere with, in the classical limit $\lambda_p\to 0$, an extra `normal' direction $\theta'$ for the sphere embedded in $\R^3$. This time the dimension jump means that the radial-time sector  matches to the closed (positively curved) 4D FLRW model. For the black hole model, the dimension jump means we land on radial and time behaviour matching the 5D Tangherlini black hole\cite{Tan} when we use the fuzzy sphere. Here the $\beta(r)=1-r_H/r$ factor in the familiar Schwarzschild  metric case for horizon radius $r_H$ is now a factor $\beta(r)=1-r_H^2/r^2$. This gets asymptotically flat faster than the Schwarzschild case and the effective gravity in the Newtonian limit is an inverse cubic force law. Finally, for the $\R^2\times\Z_n$ black hole, we have $\beta=-r_H/r$ which is as for a usual black hole but without the constant term. This therefore approximates the metric {\em inside} a Schwarzschild black hole of infinite mass (so that the missing 1 is negligible). We also cover the case $\beta(r)=r_H/r$ of interest in its own right. These models have no horizon but naked singularities. We describe the $\Z_n\to S^1$ limit where $S^1$ retains a 2D noncommutative differential structure, and the classical projection to the usual calculus on $S^1$ where the metric is no longer Ricci flat, i.e. this is a purely quantum-geometric solution of the vacuum Einstein equations. 

Our noncommutative models are theoretical and we are not aware of an immediate application, but  they do indicate an unusual phenomenon  which has a purely quantum origin in an extra `normal direction' $\theta'$ required for an associative differential calculus in our examples. We also begin to explore some of physics in our noncommutative backgrounds. In principle, one could use a new framework \cite{Beg:geo,BegMa:geo} of quantum geodesics, but the full formulation of this is rather involved and we propose instead a direct approach starting with the  Klein-Gordon equation in the noncommutative background. In Section~\ref{secqmbh}, we introduce the notion of a Schroedinger-like equation for an effective quantum theory relative to an exact solution in the same manner as usual quantum mechanics for a free particle can be obtained as a nonrelativistic limit of the Klein-Gordon equations for solutions of the form $e^{-\imath m t}\psi$ with $\psi$ slowly varying. The novel feature  will be to replace $e^{-\imath mt}$ by an exact reference solution of the Klein Gordon equation, and we explain first how this looks for a classical Schwarzschild black hole. This appears to be rather different from well-known methods of quantum field theory on a curved background \cite{Birrel, ParTom, MukWin} but fits with the general idea that a quantum geodesic flow is actually a Schroedinger-like evolution. 

The paper starts with some preliminaries in Section~\ref{secpre}, where we introduce the key points of the formalism for quantum Riemannian geometry from \cite{BegMa} by way of the quantum metric and connection for the fuzzy sphere from \cite{LirMa}, and investigate the classical limit of the latter.  We use dot or $\del_t$ for time derivatives and prime or $\del_r$ for radial derivatives (while $\del_i,\del_\pm$ will be noncommutative angular derivatives). We sum over repeated indices. $\tens_s$ will denote the symmetric tensor product, where we add the two sides flipped, and we work in units where $\hbar=c=1$.  Numerical computations were done with Mathematica. The paper concludes with some brief remarks about further directions.

\section{Recap of the fuzzy sphere and its classical limit}\label{secpre} 

The fuzzy sphere $A=\C_\lambda[S^2]$ in the sense of \cite{Hoo,BegMa,LirMa} just means the angular momentum enveloping algebra $U(su_2)$ with an additional relation giving a fixed value of the quadratic Casimir. This is the standard coadjoint quantisation of the unit sphere with its Kirillov-Kostant bracket known since the 1970s, and in our conventions takes the form
\[ [x_i,x_j]=2\imath\lambda_p \eps_{ijk}x_k,\quad  \sum_i x_i^2=1-\lambda_p^2\]
where $\lambda_p$ is a real dimensionless parameter, supposed to be of order the Planck scale relative to the actual sphere size. The conventions are chosen so that the standard spin $j$ representation descends to a representation of the the fuzzy sphere if $\lambda_p=1/(2j+1)$, but we are not restricted to these discrete values. 

The more novel ingredient in \cite{BegMa} is a 
rotationally invariant differential calculus in the sense of an exterior algebra $(\Omega,\extd)$ given by central basic 1-forms $s^i\in \Omega^1$ and exterior derivative
\[ \extd x_i = \epsilon_{ijk}x_js^k,\quad \extd s^i=-{1\over 2}\eps_{ijk}s^j\wedge s^k,\]
with associated partial derivatives defined by $\extd f(x)=(\del_i f)s^i$ in this basis (they act in the same way as orbital angular momentum). The $s^i$ are preferable as they graded commute with everything, but they can be recovered in terms of the $\extd x_i$ by\cite{LirMa}
\[ s^i={1\over 1-\lambda_p^2}(x^i\theta'+\eps_{ijk}\extd x_j x_k);\quad \theta'=x_is^i={x_i\extd x_i\over 2\imath\lambda_p}\]
There is also a $*$-operation with $x_i^*=x_i$ and $s^i{}^*=s^i$. (Then $*$ commutes with $\extd$ and $\theta'{}^*=\theta'$.)

A metric on the fuzzy sphere from the point of view of quantum Riemannian geometry means $g\in \Omega^1\tens_A\Omega^1$ subject to certain conditions, and is shown in \cite{BegMa}  to be necessarily of the form
\[ g= g_{ij}s^i\tens s^j\]
for a real symmetric matrix $g_{ij}$. Here $g$, in order to have a bimodule inverse, needs to be central and this forces the $g_{ij}$ to be constants. Quantum symmetry in the sense $\wedge(g)=0$ requires the matrix to be symmetric and reality in the sense ${\rm flip}(*\tens*)(g)=g$ then requires $g_{ij}$ to be real-valued. We also need nondegeneracy in the sense of a bimodule map $(\ ,\ ):\Omega^1\tens_A\Omega^1\to A$ inverting $g$ in the obvious way. Here `bimodule map' means commuting with the product by elements of $a$ from either side, i.e. fully tensorial from either side. In our case this is $(s^i,s^j)=g^{ij}$, the inverse matrix to $g_{ij}$. The rotationally invariant `round metric' is $g_{ij}=\delta_{ij}$ or $g=s^i\tens s^i$  (sum over $i$ understood). 

The new result in \cite{LirMa} was to find a quantum Levi-Civita connection $\nabla:\Omega^1\to \Omega^1\tens_A\Omega^1$ in the sense of torsion free and metric compatible. Here, if $X:\Omega^1\to A$ is a right module map or `right vector field' then $\nabla_X=(X\tens\id)\nabla$ is a covariant derivative on $\Omega^1$. The associated left Leibniz rule is
\[ \nabla(a \omega)=\extd a\tens\omega+ a(\nabla\omega)\]
for all $a\in A,\omega\in \Omega^1$. We define the torsion tensor and Riemann curvature tensor for any connection as maps\cite{BegMa}
\[ T_\nabla:\Omega^1\to \Omega^2,\quad T_\nabla=\wedge\nabla-\extd,\quad R_\nabla:\Omega^1\to \Omega^2\tens_A\Omega^1,\quad R_\nabla=(\extd\tens\id-\id\wedge\nabla)\nabla.\]
However, as we can also multiply 1-forms by algebra elements from the right, we need another Leibniz rule\cite{DVM}
\[ \nabla(\omega a)=(\nabla\omega)a+\sigma(\omega\tens\extd a)\]
for some bimodule map $\sigma:\Omega^1\tens_A\Omega^1\to \Omega^1\tens_A\Omega^1$.  The generalised braiding map $\sigma$ is not additional data as it is determined by the above if it exists. Connections where it exists are called `bimodule connections' and we will focus on these. We then define the metric compatibility tensor by 
\[  \nabla g:= (\nabla\tens {\rm id} + (\sigma\tens\id)(\id\tens\nabla))g.\]
The connection is quantum Levi-Civita if $T_\nabla=\nabla g=0$. We also require a reality condition
\[ \sigma\circ{\rm flip}(*\tens*)\circ\nabla=\nabla\circ *.\]
Finally, for physics we need a Ricci tensor and the working definition\cite{BegMa} is to suppose a bimodule map $i:\Omega^2\to \Omega^1\tens_A\Omega^1$ and define ${\rm Ricci}$ by a trace of $(i\tens\id)R_\nabla:\Omega^1\to \Omega^1\tens_A\Omega^1\tens_A\Omega^1$. This can be done explicitly via the metric and its inverse  to make the trace between the input and the {\em first} factor,
\begin{equation*} {\rm Ricci} = ((\ ,\ ) \otimes \id)(\id\otimes i \otimes \id)(\id\otimes R_\nabla)g.\quad 
	\end{equation*}
These natural definitions mean, however, that {\rm Ricci} as it comes out from quantum Riemannian geometry is $-{1\over 2}$ of the usual Ricci in the classical case. The Ricci scalar  $S=(\ ,\ ){\rm Ricci}$ is also $-{1\over 2}$ of the usual one.  

All of this can be solved for the fuzzy sphere under the assumption that the coefficients are constant in the $s^i$ basis, giving\cite{LirMa}
\[ \nabla s^i=-{1\over 2}\Gamma^i{}_{jk}s^j\tens s^k,\quad \Gamma^i{}_{jk}=g^{il}(2\eps_{lkm}g_{mj}+{\rm Tr}(g)\eps_{ljk}).\]
Moreover, as classically, we can just take the map $i$ to be the antisymmetric lift, so 
\[ i(s^i\wedge s^j)={1\over 2}(s^i\tens s^j-s^j\tens s^i).\]
The resulting Ricci curvature on the fuzzy sphere are in \cite{LirMa} but in the round metric case one has
\[ \nabla s^i=-{1\over 2}\eps_{ijk}s^j\tens s^k,\quad {\rm Ricci}=-{1\over 4} g,\quad S=-{3\over 4}.\]
The curvatures here are not the values you might have expected for a unit sphere even allowing for our conventions.  Nor does the Einstein tensor (at least, if defined in the usual way) vanish as would be the case for a classical 2-manifold. 

To understand this last point better, which is the modest new result of this preliminary section, we look more carefully at the classical limit $\lambda_p\to 0$. By the Leibniz rule and the values above, we have for the round metric
\begin{align*}\nabla(\theta')&=x_i\nabla s^i+\extd x_i\tens s^i=-{1\over 2}\eps_{ijk}x_is^j\tens s^k+ \eps_{ijk}x_j s^k\tens s^i={1\over 2}\eps_{ijk}x_is^j\tens s^k\\
&={1\over 2(1-\lambda_p^2)^2}\eps_{ijk}(x_j\theta'+\eps_{jmn}(\extd x_m) x_n)\tens(x_k\theta'+\eps_{kab}(\extd x_a) x_b)\\
&={1\over 2}\eps_{ijk}\extd x^i\tens (\extd x^j)x^k + O(\lambda_p).
\end{align*}
This means that we cannot just set $\theta'=0$ in the classical limit for the given quantum geometry. Meanwhile, we write the commutation relations of the calculus as
\[ [\theta', x_i]=2\imath\lambda_p \extd x_i, \quad x_i\extd x_i=2\imath\lambda_p \theta',\quad [x_i,\extd x_j]=2\imath\lambda_p(\delta_{ij}\theta'-{x_j\over 1-\lambda_p^2} (x_i\theta'+\eps_{imn}(\extd x_m)x_n))\]
from which we see that the calculus is commutative and $x_i\extd x_i=0$ in the classical limit $\lambda_p\to 0$ as expected for
the unit sphere, but $\theta'$ itself does not need to vanish, and we have seen that it cannot if we want to have a limit for $\nabla$. Rather, we consider the classical limit as the classical sphere plus a single remnant $\theta'$ which graded-commutes with everything and (in the classical limit) does not arise from functions and differentials on the sphere. Indeed, this limit is not a strict differential calculus but a generalised one for this reason, but there is no such problem in the quantum case, where 
\[ \theta'={1\over 2\imath\lambda_p}x_i\extd x_i\]
shows its origin as `normal' to the sphere as embedded in $\R^3$. We now note that the round metric has the limit
\begin{align*} g&=s^i\tens s^i= (x_i\theta'+\eps_{imn}(\extd x_m) x_n)\tens (x_i\theta'+\eps_{iab}(\extd x_a) x_b)\\
&=(1-\lambda_p^2)\theta'\tens\theta'+\eps_{imn} x_i\theta'\tens_s(\extd x_m)x_n+(\delta_{ma}\delta_{nb}-\delta_{mb}\delta_{na})(\extd x_m)x_n\tens(\extd x_a)x_b\\
&=\theta'\tens\theta'+\extd x_i\tens\extd x_i+ O(\lambda_p)\end{align*}
since the calculus is commutative to $O(\lambda_p)$.  Thus we see that the rotationally invariant `round'  metric actually has an extra direction required by the calculus. We can recover the completely classical $S^2$ by the limit $\lambda_p\to 0$ {\em and} projecting $\theta'=0$, but traces taken for the Ricci curvature before we do this will remember the extra `normal direction' and not map onto the classical values.

\section{Expanding fuzzy sphere FLRW model}\label{secFLRW}

Here we work with the coordinate algebra $A=C^\infty(\R) \tens  \C_\lambda[S^2]$ where the $\R$ has a classical time $t$ variable with classical $\extd t$ graded commuting with $t,\extd t$ and with the generators $x_i,s^i$ of the exterior algebra of the fuzzy sphere.  
We first consider a general metric of the form
\begin{equation}
	\label{eq:general_metric}
	g = \beta \extd t \tens \extd t + n_i (\extd t \tens s^i + s^i \tens \extd t) + g_{ij} s^i\tens s^j
\end{equation}
where $g_{ij}$ is a symmetric $3\times 3$ matrix of coefficients and $\beta, n_i$ further coefficients, {\em a priori} all  valued in $A$. Centrality of the metric, however, then forces the $n_i = 0$ and the remaining coefficients to be in the centre of $C_\lambda[S^2]$, which is trivial. Hence $g_{ij},\beta$ are functions only of the time $t$. The reality condition for quantum metrics forces them to be real-valued. 

Next,  a general QLC for the calculus has the form
\begin{align}
	\label{eq:mostgeneral_QLC}
	\nabla s^i =& -\frac{1}{2}\Gamma^i{}_{jk}s^j\tens s^k + \gamma^i{}_j s^j\tens_s \extd t + \tau^i \extd t \tens \extd t\\
	\nabla \extd t =& \mu_{jk}s^j\tens s^k + \eta_j s^j\tens_s \extd t + \Gamma \extd t \tens \extd t
\end{align}
again with $\Gamma^i{}_{jk},\gamma^i{}_j, \tau^i, \Gamma, \eta_j,\mu_{jk}\in A$. However, given that the spatial metric $g_{ij}$ are functions only of $t$, it is natural to assume this also for the spatial Christoffel symbols $\Gamma^i_{jk}$ just as is done for the fuzzy sphere alone in\cite{LirMa}. In this case, compatibility of $\nabla$ with the relations of commutativity of $\extd t,s^i$ with $t,x_j$ and the natural assumption that the associated braiding $\sigma$ has the classical `flip' form when one of the arguments is $\extd t$, conditions needed for a bimodule connection,  require that $\gamma^i_j, \tau^i, \Gamma, \eta_j,\mu_{jk}$ are also function of time alone.

The non trivial conditions for $\nabla$ to be torsion-free are
\begin{equation}\label{flrwtor} \frac{1}{2}(-\Gamma^i{}_{jk} + \epsilon^i{}_{jk}  ) s^j\tens s^k = 0, \quad \mu_{jk} = \mu_{kj},\end{equation}
since $\extd (\extd t) = 0.$ The conditions  $\nabla g = 0$ for metric compatibility then produces
\begin{align*}
	\extd t \tens s^i\tens s^j &: \dot{g}_{ij} + g_{il}\gamma^l{}_j + g_{lj}\gamma^l{}_i = 0, \\
	\extd t \tens \extd t\tens \extd t &: \dot{\beta} + 2 \beta\Gamma = 0, \\
	s^l\tens s^m \tens s^j  &:  -\frac{g_{ij}}{2}\Gamma^i{}_{lm} - \frac{g_{in}}{2}\Gamma^n{}_{pj}\sigma^{ip}{}_{lm} = 0,\\
	s^l\tens \extd t \tens s^j  &: g_{ij}\gamma^i{}_l + \beta \mu_{lj} = 0, \\
	\extd t\tens \extd t \tens s^j  &: g_{ij}\tau^i + \beta\eta_j = 0, \\
	s^n\tens s^p\tens \extd t  &:  g_{ij} \gamma^j{}_l \sigma^{il}{}_{np} + \beta\mu_{np} = 0,\\
	\extd t\tens s^i \tens \extd t  &: g_{ij}\tau^j + \beta\eta_i= 0, \\	
	s^m\tens \extd t \tens \extd t  &:  2\beta\eta_m= 0.
\end{align*}
It is clear from third of these and the first of (\ref{flrwtor}) that $\Gamma^{i}_{jk}$ is indeed the Christoffel symbol for the fuzzy sphere QLC as solved uniquely in the $*$-preserving case with constant coefficients  in \cite{LirMa}. Also, the last equation implies that $ \eta_{m} = 0$ and,  using this together with the fifth or seventh equation, we get $\tau^i = 0$. The second equation makes $\Gamma = -\dot\beta/(2\beta)$.  Using $g_{ik}\gamma^{k}{}_{j} = \gamma_{ij}$ and the symmetry of $g_{ij}$ in the first equation  we get the value of $\gamma^{i}{}_{j}$, then  this together with the 4th equation gives  $\mu_{ij}$, resulting in
\begin{equation}
	\gamma^{i}{}_{j} = -{1\over2}\dot{g}_{jk}g^{ik}, \quad \mu_{ij} = {\dot{g}_{ij} \over 2\beta}.
\end{equation}
Note that $\mu_{ij}$ is proportional to the time derivative of the metric, which implies that it is also symmetric if the metric is, solving the second half of (\ref{flrwtor}).  Because $\gamma^i{}_j, \mu_{ij}$ just depend on real functions, they are also real-valued functions.   This leads to  a reasonably canonical QLC. 

\begin{theorem} Up to a reparametrisation of $t$, a quantum metric on the algebra $C^\infty(\R)\tens C_\lambda[S^2]$ has to have the form \[ g = -\extd t \tens \extd t + g_{ij} s^i\tens s^j , \]  where $g_{ij}$ is a time-dependent real $3\times 3$ symmetric matrix. Moreover, this admits  a canonical $*$-preserving QLC
\begin{align*}
	\nabla \extd t = -\frac{1}{2} \dot{g}_{ij}s^i\tens s^j, \quad \nabla s^i = -\frac{1}{2}\Gamma^{i}{}_{jk} s^i\tens s^j - \frac{1}{2}g^{ki}\dot g_{jk} s^j\tens_s \extd t,
\end{align*}
where $\Gamma_{ijk} = 2\epsilon_{ikm}g_{mj} +{\rm Tr}(g)\epsilon_{ijk}$ as for the fuzzy sphere in \cite{LirMa}. The associated Ricci scalar and Laplacian are
\begin{align*}
2S &= - g^{ij}\ddot{g}_{ij} -Tr(g) + \frac{1}{2}(Tr(g))^2 - \delta_{ij} - \frac{1}{4}\left(  g^{ml}g^{ij}\dot{g}_{ml}\dot{g}_{ij}  + g^{kl}g^{mn}\dot{g}_{nk}\dot{g}_{lm} \right),\\ 
 \Delta f &= \left( g^{ji}\del_{j}\del_{i} - \frac{1}{2}\left( g^{ij}\dot{g}_{ij} \right)\del_{t} -\del^2_{t}\right)f. \end{align*}
\end{theorem}
\proof
The analysis for the metric was done above and we were forced by the requirement for the metric to be central (in order to be invertible) to $n_i=0$ and $\beta(t)$, $g_{ij}(t)$ in (\ref{eq:general_metric}). We add the $*$-reality of the metric in the form ${\rm flip}(*\tens *)g=g$ to find $\beta$ and $g_{ij}$ real. Quantum symmetry also requires the latter to be symmetric. By a change of $t$ variable, we can generically assume $\beta=-1$,  but we do not need to do this. 

Now substituting the obtained values so far in the analysis of the general form of the QLC (\ref{eq:mostgeneral_QLC}), we have  the connection
\begin{equation}
	\label{eq:general_QLC}
	\nabla \extd t = \frac{1}{2\beta} \dot{g}_{ij}s^i\tens s^j - \frac{1}{2}\frac{\dot{\beta}}{\beta} \extd t \tens \extd t; \quad \nabla s^i = -\frac{1}{2}\Gamma^{i}{}_{jk} s^j\tens s^k - \frac{1}{2}g^{ki}\dot g_{jk} s^j\tens_s \extd t
\end{equation}
for some unknown $\Gamma^i{}_{jk}(t)$, where we assumed that this does not depend on fuzzy sphere variables (which is reasonable given that the metric can not). The requirement of being $*$-preserving yields
\begin{equation}\label{flrwstar} \dot g_{jk}(s^j\tens s^k -\sigma(s^k\tens s^j)) = 0, \quad  \Gamma^i{}_{jk}s^j\tens s^k - \overline{\Gamma}^i{}_{kj}\sigma(s^k\tens s^j) = 0.\end{equation}
 with the second of these the same as for the fuzzy sphere in \cite{LirMa} at each fixed time.  Here we used $\extd t^* = \extd t$. Thus all the equations for $\Gamma^i{}_{jk}$ are the same at in that paper and hence there is a unique solution for it in terms of $g_{ij}(t)$,  as stated,  under the assumption of no fuzzy sphere dependence. In this case, we know from \cite{LirMa} that $\sigma=$flip on $s^j\tens s^k$ and hence the first of (\ref{flrwstar}) is empty, as is the 6th of the metric compatibility equations in our previous analysis. The rest of the $*$-preserving conditions require  $\Gamma, \eta_i, \gamma^i{}_j$ to be real-valued functions, which already holds as we have solved for them. 

The curvature for the connection (\ref{eq:general_QLC}) is
\begin{align*}
	R_\nabla \extd t =& \left( {1\over 2\beta}  \ddot g_{ij} - {\dot \beta\over 4\beta^2} \dot g_{ij} - {1\over 4\beta} g^{ml}g_{il}g_{jm} \right) \extd t \wedge s^i \tens s^j \\
+& {1\over 4\beta} \left( -\dot g_{lk}\epsilon^{l}{}_{ij} + \dot g_{il} \Gamma^{l}{}_{jk}  \right) s^i \wedge s^j \tens s^k 
+ {1\over 4\beta} g^{lm}\dot g_{jl} \dot g_{im} s^i \wedge s^j \tens \extd t \\
	R_\nabla s^i =& \left(  {1\over 4}\Gamma^{i}{}_{jl}g^{ml}\dot g_{km} - {1\over 4}\Gamma^{l}{}_{jk}g^{mi}\dot g_{lm} \right) \extd t \wedge s^j \tens s^k  +  \left(  {1\over 4}g^{li}\dot g_{ml}\epsilon^{m}{}_{jk} - {1\over 4}\Gamma^{i}{}_{jl}g^{ml}\dot g_{km} \right)s^j \wedge s^k \tens \extd t \\
	+&  \left(  {1\over 4}\Gamma^{i}{}_{ml}\epsilon^{m}{}_{jk} - {1\over 4} \Gamma^{i}{}_{jm} \Gamma^{m}{}_{kl} + {1\over 4\beta} g^{mi}\dot g_{jm} \dot g_{kl} \right) s^j \wedge s^k \tens s^l \\
	+&  \left(  -{1\over 2}(\dot g^{ki} \dot g_{jk} + g^{ki}\ddot g_{jk}) + {\dot \beta\over 4\beta}g^{ki}\dot g_{jk} - {1\over 4}g^{ki}g^{ml}\dot g_{lk}\dot g_{jm} \right)\extd t \wedge s^j \tens \extd t,
\end{align*}

For this connection we have the Ricci tensor as follows
\begin{align*}
	2 {\rm Ricci} &= \left( \frac{1}{2\beta}\left(  \ddot{g}_{ij} -\frac{\dot{\beta}}{2\beta}\dot{g}_{ij} - g^{kl}\dot{g}_{il} \dot{g}_{jk}  \right) + \frac{1}{2} \Gamma^{l}{}_{mj}\epsilon^{m}{}_{li} -\frac{1}{4}\Gamma^{l}{}_{lm}\Gamma^{m}{}_{ij} + \frac{1}{4\beta} g^{ml}\dot{g}_{ml}\dot{g}_{ij} \right. \\
	&+ \left.   \frac{1}{4}\Gamma^{l}{}_{im}\Gamma^{m}{}_{lj} - \frac{1}{4\beta}g^{ml}\dot{g}_{im}{}\dot{g}_{lj} \right) s^i\tens s^j \\
	&-\left(  -\frac{1}{2}\left( \dot{g}^{kl}\dot{g}_{lk} + g^{ij}\ddot{g}_{ij} \right) + \frac{\dot{\beta}}{4\beta} g^{ij}\dot{g}_{ij} - \frac{1}{4}g^{kl}g^{mn}\dot{g}_{nk}\dot{g}_{ml} \right) \extd t \tens \extd t \\
	&+  \left(  \frac{1}{2}g^{nl}\dot{g}_{mn}\epsilon^{m}{}_{il} - \frac{1}{4}\Gamma^{l}{}_{lm}g^{nm}\dot{g}_{in} + \frac{1}{4}\Gamma^{l}{}_{im}g^{nm}\dot{g}_{ln} \right) s^i\tens \extd t \\
	&+  \left( - \frac{1}{4}\Gamma^{l}{}_{lm} g^{nm}\dot{g}_{in}  +\frac{1}{4}g^{nl}\dot{g}_{mn}\Gamma^{m}{}_{li} \right) \extd t \tens s^i
\end{align*}

Now taking $\beta=-1$, and the explicit value of  $\Gamma^{i}{}_{jk}$, the Ricci tensor follows as
\begin{align*}
	2 {\rm Ricci} &= \left(   -\frac{\ddot{g}_{ij}}{2}  + \frac{1}{2} g^{kl}\dot{g}_{il} \dot{g}_{jk}  - \frac{1}{4} g^{ml}\dot{g}_{ml}\dot{g}_{ij}  -g_{ij} - \delta_{ij} + \frac{1}{2}{\rm Tr}(g)g_{ij}   \right) s^i\tens s^j \\
	&-\left(  -\frac{1}{2}\left( \dot{g}^{kl}\dot{g}_{lk} + g^{ij}\ddot{g}_{ij} \right) - \frac{1}{4}g^{kl}g^{mn}\dot{g}_{nk}\dot{g}_{ml} \right) \extd t \tens \extd t \\
	&+  \left(  \frac{1}{2}g^{nl}\dot{g}_{mn}\epsilon^{m}{}_{il}  -\frac{1}{2}\epsilon^{kl}{}_{m}g^{nm}g_{kl}\dot{g}_{in} + \frac{1}{4}(2\epsilon^{kl}{}_{m}g_{ik}+{\rm Tr}(g)\epsilon^{l}{}_{im}  )g^{nm}\dot{g}_{ln} \right) s^i\tens \extd t \\
	&+  \left( - \frac{1}{2}\epsilon^{kl}{}_{m}g_{kl} g^{nm}\dot{g}_{in}  -\frac{1}{4}g^{nl}\dot{g}_{mn}(2\epsilon^{km}{}_{i}g_{kl} + {\rm Tr}(g) \epsilon^{m}{}_{li} ) \right) \extd t \tens s^i
\end{align*}
Making the contraction with the inverse metric  we recover the required Ricci scalar.

The Laplacian for a function $f = f(t,x^i)$ follows as 
\[ \Delta f = (,) \nabla(\extd f) = (,) \nabla(\del_i f s^i + \dot{ f }\extd t) = g^{ij}\del_{i}\del_{j}f - \del^2_tf - {1\over 2} g^{ij}\dot{g}_{ij}\del_tf - {1\over 2}g^{jk}\Gamma^i{}_{jk}\del_if,\]
where the last term vanish when we take into account the explicit form of $\Gamma^i{}_{jk}$, recovering the required Laplacian.
\endproof

The QLC here is unique under the reasonably assumption is in \cite{LirMa} that the $\Gamma^i{}_{jk}$ are constant on the fuzzy sphere, given that the $g_{ij}$ have to be.  The theorem applies somewhat generally but now we take the expanding round metric $g_{ij} = R^2(t)\delta_{ij}$ for the spatial part, so the metric, non-zero inverse metric entries, QLC, curvature and Laplacian are
\begin{align} g &= - \extd t \tens \extd t + R^2(t) s^i \tens s^i,\quad (\extd t,\extd t)=-1 ,\quad  (s^i,s^j) = {\delta^{ij} \over R^2},\\
\nabla \extd t &= - R\dot R s^i \tens s^i; \quad \nabla s^i = -\frac{1}{2} \epsilon^{i}{}_{jk}s^j \tens s^k - \frac{\dot{R}}{R} s^i \tens_s \extd t, \\
 R_\nabla \extd t &= -R\ddot R \extd t \wedge s^i\tens s^i, \\
 R_\nabla s^i &= \left(  {1\over 4} \epsilon^{pi}{}_{n}\epsilon_{pkm} - \dot R^2 \delta^{i}{}_{m}\delta_{nk} \right) s^m \wedge s^n \tens s^k + {\ddot R \over R}\extd t \wedge s^i \tens \extd t,\\
 {\rm Ricci} &= -(  \dot R^2 + {1\over 2}R\ddot R +\frac{1}{4})s^i \tens s^i + {3\over 2}{\ddot R \over R} \extd t \tens \extd t,\quad S =-3\left(  \frac{\dot{R}^2}{R^2} + \frac{\ddot{R}}{R} +\frac{1}{4R^2} \right),\\
\Delta &= \frac{1}{R^2}\sum_i\del^2_i - 3\frac{\dot{R}}{R}\del_{t} - \del^2_t.\end{align}

Also of interest is the Einstein tensor and, in the absence of a general theory, we assume as in \cite{ArgMa} the `naive definition' ${\rm Eins}={\rm Ricci}-{S\over 2}g$, which works out as. 
\begin{equation}\label{flrwEins} {\rm Eins} = \left(  \ddot{R} + \frac{1}{2}\dot{R}^2 +\frac{1}{8}\right)s^i \tens s^i - \frac{3}{2}\left(   \frac{1}{4R^2} +\frac{\dot{R}^2}{R^2} \right) \extd t \tens \extd t \end{equation}
and is justified by checking that
\begin{equation}\label{einscons} \nabla\cdot{\rm Eins}=0.\end{equation}
Here, if we have any tensor for the form $T = f \extd t \tens \extd t + p R^2 s^i \tens s^i$, then the divergence is
\[ \nabla \cdot T =( (\ ,\ )\tens\id)\nabla T = -\left(  \dot{ f } +3( f+p )\frac{\dot{R}}{R} \right)\extd t + \del_{i} p s^i \]
and we use this now for the particular form of the Einstein tensor to establish (\ref{einscons}). We also assume this form of $T$ for the  energy-momentum tensor of dust with pressure $p$ and density $f$, in which case the continuity equation $\nabla\cdot T=0$ for $p$ a function only of $t$ is 
\[ \dot f+3(f+p)\frac{\dot R}{R}=0\]
as usual, and Einstein's equation ${\rm Eins}+4\pi G T=0$ in our curvature conventions is
\[  4\pi G f = \frac{3}{2}\left( \frac{\dot{R}^2}{R^2} + \frac{1}{4R^2} \right),\quad 4\pi G p = -\frac{\ddot{R}}{R} - \frac{1}{2}\frac{\dot{R}^2}{R^2} -\frac{1}{8R^2}=  -\frac{\ddot{R}}{R} - {4\pi G\over 3}f  \]
These  are identical to the classical FLRW equations, see e.g.\cite[Chap. 8]{Car}, for a 4D closed universe with curvature constant $\kappa = 1/(4R_0^2)$ in the classical FLRW metric
\[ -\extd t\tens\extd t+ R(t)^2\left({1\over r^2(1-\kappa r^2)}\extd r\tens\extd r+ g_{S^2}\right),\]
where $g_{S^2}$ is the metric on a unit sphere, $R_0$ is a normalisation constant with dimension of length, and we have adapted $R(t)$ to include $r$ in order to match our conventions. 

\section{Black hole with the fuzzy sphere}\label{secBHfuz}
We assume a similar framework as in the previous section, but now with a 4D metric of a static form in polar coordinates. Thus, we add a radial variable $r$ with differential $\extd r$ and consider the Schwarzschild-like metric
\begin{equation}\label{genSchlikeg} g = -\beta(r)\extd t \tens \extd t + H(r) \extd r \tens \extd r + r^2 g_{ij}s^i\tens s^j. 
\end{equation}
The algebra of functions here is $A=C^\infty(\R\times\R_{>0})\tens \C_\lambda[S^2]$ with classical variables and differentials $t,r,\extd r, \extd t$ for the $\R\times\R_{>0}$ part (so these graded commute among themselves and with the functions and forms on the fuzzy sphere). The coefficients $g_{ij}$ define the metric on the fuzzy sphere, and centrality and reality of the metric dictates that these are constant real values. Thus, $g_{ij}$ is a real symmetric invertible $3\times 3$ matrix (it should also be positive definite for the expected signature) and $\beta(r), H(r)$ are real-valued functions. 

We start with the general form of connection on the tensor product calculus,
\begin{align*}
	\label{eq:general SCF}
\nabla s^i &= -{1\over 2}\Gamma^i_{jk}s^j\tens s^k + \alpha^i \extd t \tens \extd t + \gamma^i \extd r \tens \extd r+ \Delta^i \extd r \tens_s \extd t + \eta^i{}_j \extd t \tens_s s^j + \tau^i{}_j \extd r \tens_s s^j,
\\
\nabla \extd t &= a_{ij}s^i\tens s^j + b \extd t \tens \extd t + c \extd r \tens \extd r+ d \extd r \tens_s \extd t + e_j \extd r \tens_s s^j + f_j \extd t \tens_s s^j,
\\
\nabla \extd r&= h_{ij}s^i\tens s^j + \theta \extd t \tens \extd t + R \extd r \tens \extd r+ \phi \extd r \tens_s \extd t + \nu_j \extd t \tens_s s^j + \psi_j \extd r \tens_s s^j.
\end{align*}
 Assuming that $\sigma(\extd t\tens), \sigma(\tens\extd t),\sigma(\extd r\tens), \sigma(\tens\extd r)$ are the flip on the 1-forms $\extd r, \extd t, s^i$ and the natural restrictions needed for a bimodule connection, one finds that all the coefficients are functions of $t$ and $r$ alone (constant on the fuzzy sphere).

The torsion freeness conditions for $\nabla \extd t, \nabla\extd r$ and $\nabla s^i$ are 
\begin{equation}
	a_{ij} = a_{ji}, \quad h_{ij} = h_{ji}, \quad \Gamma^{i}{}_{jk} - \Gamma^{i}{}_{kj} + 2 \epsilon^{i}{}_{jk} = 0,
\end{equation}
respectively, and the conditions needed for the compatibility with the metric are
\begin{align*}
	\extd r\tens \extd t\tens \extd t &: \del_r\beta + 2\beta d = 0, \\
	\extd r^{\tens 3} &: \del_r H + 2H R = 0, \\
	\extd r\tens s^l\tens s^j &: 2rg_{lj} + r^2 g_{ij}\tau^{i}{}_{l} + r^2 g_{lm}\tau^{m}{}_{j} = 0, \\
	s^m \tens s^n \tens \extd t &: -\beta_{mn} + r^2g_{ij} \eta^{j}{}_{l}  \sigma^{il}{}_{mn} = 0, \\
	\extd t^{\tens 3} &: -2\beta b = 0, \\
	\extd r\tens \extd r\tens \extd t / \extd r\tens \extd t\tens \extd r &: -\beta c + H \phi = 0, \\
	\extd t\tens \extd r\tens \extd r &:  2H\phi  = 0, \\
	s^i\tens \extd t\tens \extd r / s^i\tens \extd r\tens \extd t &: -\beta e_i + H\nu_i = 0, \\
	\extd r\tens s^j\tens \extd t / \extd r\tens \extd t \tens s^j &: -\beta e_j + r^2 g_{ij}\Delta^i = 0, \\
	s^i \tens \extd t\tens \extd t &: -2\beta f_j = 0, \\
	\extd t\tens \extd t \tens s^j / \extd t\tens s^j \tens \extd t &: -\beta f_j + r^2g_{ij}\alpha^i = 0, \end{align*}\begin{align*} 
	s^i\tens \extd t\tens s^j &: -\beta a_{ij} + r^2 g_{lj}\eta^{l}{}_{i} = 0, \\
	\extd t\tens \extd r\tens \extd t / \extd t\tens \extd t\tens \extd r &: -\beta d + H\theta = 0, \\
	\extd r\tens \extd t\tens \extd t &: \del_r\beta + 2\beta d = 0, \\
	s^m\tens s^n\tens \extd r  &: Hh_{mn} + r^2 g_{ij} \tau^{j}{}_{l}\sigma^{il}{}_{mn} = 0, \\
	\extd t\tens \extd r \tens s^i /\extd t\tens s^i\tens \extd r &: H\nu_{i} + r^2g_{ij}\Delta^j = 0, \\
	\extd r\tens \extd r\tens s^i/\extd r\tens s^i\tens \extd r &: H\psi_i + r^2g_{ij}\gamma^j = 0, \\	
	s^i\tens \extd r\tens \extd r &: 2H\psi_i = 0, \\
	s^i\tens \extd r\tens s^j &: Hh_{ij} + r^2 g_{lj}\tau^{l}{}_{i} = 0, \\
	s^p\tens s^q\tens s^m &: g_{lm}\Gamma^{l}{}_{pq} + g_{ij}\Gamma^{j}{}_{lm}\sigma^{il}{}_{pq} = 0, \\
	\extd t\tens s^i\tens s^j &: g_{lj}\eta^{l}{}_{i} + g_{il}\eta^{l}{}_{j}= 0. 
\end{align*}
We immediately note that $b=\phi=f_j=\psi_i=0$ for the 5th, 7th, 10th, and 18th equations respectively.  In this case, we have that  $\alpha^i = \phi = \gamma^i = 0$ by the  11th, 6th and 17th equations respectively. Also, solving simultaneously the  8th, 9th, 16th equations, we obtain $\Delta^i = e_i = \nu_i = 0$. The value of $d$ and $R$ is deduced for the 1st and 2nd equations respectively, while $\theta$ comes from 13th and 1st equations, with result\begin{equation}
	d = -{\del_r\beta\over 2\beta}, \quad R = -{ \del_r H \over 2H}, \quad \theta = - { \del_r\beta \over 2H}.
\end{equation}
The 3rd equations together with the symmetry of $g_{ij}$ lead to $\tau^{i}{}_j = -{1\over r} \delta^{i}{}_{j}$. Now, we can solve the 19th equation as 
\begin{equation}
	\label{eq:h-exp}
	h_{ij} = {r\over H} g_{ij}.
\end{equation}
The 21st equation gives the condition $\eta_{ij} - \eta_{ji} = 0$, where we used $\eta^k{}_jg_{ki}=\eta_{ij}$. But the 12th equation produces  $a_{ij} = {r^2\over \beta}\eta_{ij}$ so that $a_{ij}$ is anti-symmetric, which together with the torsion freeness conditions imply that $a_{ij} = \eta^i{}_j = 0$.

\begin{theorem} The static Schwarzschild-like metric with spatial part  a fuzzy sphere, 
\[g = -\beta(r)\extd t \tens \extd t + H(r) \extd r \tens \extd r + r^2 g_{ij}s^i\tens s^j, \]
where $g_{ij}$ is a real symmetric  matrix with entries constant on the fuzzy sphere, has a canonical $*$-preserving QLC given by
\begin{align*} 
	\nabla \extd t &= -\frac{1}{2\beta} \del_r\beta \extd r \tens_s \extd t,  \quad
	\nabla \extd r = -\frac{1}{2H} \del_r H \extd r \tens \extd r + {r\over H} g_{ij} s^i \tens s^j - \frac{1}{2H}\del_r\beta \extd t \tens \extd t,\\
	\nabla s^i &= -\frac{1}{2}\Gamma^{i}{}_{jk} s^j\tens s^k - \frac{1}{r} \extd r \tens_s s^i,
\end{align*}
where $\Gamma_{ijk} = 2\epsilon_{ikm}g_{mj} +{\rm Tr}(g)\epsilon_{ijk}$ is the fuzzy sphere QLC  from \cite{LirMa}. The corresponding Ricci scalar and Laplacian are
\begin{align*}S &=  {1\over 2H\beta}\del^2_r\beta - {1\over 4 H\beta^2}(\del_r\beta)^2 - {1\over 4H^2\beta}\del_r\beta\del_rH + {3\over 2r^2} \\
&\quad+ {1\over 4rH}(3+Tr(g))\left({\del_r\beta\over \beta} - {\del_rH\over H}\right)   + {Tr(g)\over r^2H}(1-{H\over 2}) + {(Tr(g))^2\over 4r^2},\\
\Delta &= -{1\over \beta}\del^2_t  + {1\over H} \del^2_r + \left( {3\over rH } - {\del_rH\over 2H^2} + {\del_r\beta\over 2H\beta} \right)\del_r + {g^{ij}\over r^2} \del_i\del_j.\end{align*}
\end{theorem}
\proof Most of the analysis was done above. The torsion-freeness and metric compatibility conditions for Christoffel symbol $\Gamma$ of the fuzzy sphere part are the same as in \cite{LirMa} as is the second half of the $*$-preserving conditions 
\[ h_{ij} s^i\tens s^j - \overline{h}_{ji }\sigma(s^j\tens s^i) = 0, \quad  \Gamma^{i}{}_{jk} s^j\tens s^k - \overline{\Gamma}^{i}{}_{kj}\sigma(s^k\tens s^j) = 0\]
coming from $\nabla\extd r$ and $\nabla s^i$ respectively, with $(s^i)^*= s^i, \extd r^*  = \extd r$ and $\extd t^* = \extd t$. There is therefore a unique solution for $\Gamma$ under the assumption that it consists of constants according to \cite{LirMa}, and we use this solution. This has $\sigma$ the flip on the $s^i$ and hence $\Gamma$ real. In this case, the other condition for $*$-preserving requires $h_{ij}$ to be hermitian, which already holds because $h_{ij}$ is real and symmetric for (\ref{eq:h-exp}).  The 4th and 15th metric compatibility  equations  also then hold. The connection stated is then obtained by substituting into the general form of the connection. This completes the analysis for the canonical QLC. 

The curvature for this connection comes out as
\begin{align*}
	{\rm R}_\nabla s^i &= \left( {1\over4}\Gamma^{i}{}_{jk}\epsilon^{j}{}_{mn} - {1\over 4}\Gamma^{i}{}_{ml}\Gamma^{l}{}_{nk} + {1\over H}g_{nk}\delta^{i}{}_{m}\right)s^m\wedge s^n\tens s^k \\
	&\quad - {1\over 2rH}\del_r H s^i\wedge \extd r \tens \extd r + {1\over 2r}\left( \epsilon^{i}{}_{jk} - \Gamma^{i}{}_{jk} \right)s^j \wedge s^k\tens \extd r -{1 \over 2rH}\del_r\beta s^i\wedge \extd t \tens \extd t\\
	{\rm R}_\nabla \extd t &= \left( {\del^2_r\beta\over 2\beta} - \left( {\del_r\beta\over 2\beta} \right)^2 - {1\over 4\beta H} \del_r \beta \del_r H\right)  \extd t \wedge \extd r \tens \extd r + {r\over 2H\beta}\del_r\beta g_{ij} \extd t  \wedge s^i \tens s^j\\
	{\rm R}_\nabla \extd r &= - {r\over 2H^2}g_{ij}\del_r H \extd r \wedge s^i\tens s^j + {r\over 2H}(g_{ml}\Gamma^{l}{}_{nj} - g_{ij}\epsilon^{i}{}_{mn}) s^m \wedge s^n \tens s^j \\
	&\quad + \left( {1\over 4H^2} \del_r\beta \del_r H - {1\over 2H} \del^2_r \beta + {1\over 4 \beta H} (\del_r\beta)^2\right) \extd r \wedge \extd t \tens \extd t + {g_{ij}\over H} s^i\wedge s^j \tens \extd r.
\end{align*}
Taking the antisymmetric lift of products of the basic 1-forms and tracing gives the associated Ricci tensor
\begin{align*}
	4{\rm Ricci} &=  ( {\del_r^2\beta\over \beta} - \frac{3}{rH} \del_r H - {1\over 2 \beta H}\del_r\beta \del_rH - {1\over 2}\left({\del_r\beta \over \beta}\right)^2  ) \extd r \tens \extd r \\
	&\quad+{1\over H}\left( {1\over 2H}\del_r\beta\del_r H - \del_r^2\beta + {1\over 2\beta}(\del_r\beta )^2 - {3\over r} \del_r\beta \right) \extd t \tens \extd t \\
	&\quad + (  {r\over H\beta}g_{ij}\del_r\beta  - {r\over H^2}g_{ij}\del_r H  +  4{g_{ij}\over H} -2g_{ij} -2\delta_{ij} + {\rm Tr}(g)g_{ij} )s^i \tens s^j.
\end{align*}
This gives the Ricci scalar as stated. The Laplacian is also immediate from $\nabla$ and the inverse metric.  \endproof

The QLC here is unique under the reasonable assumption as in \cite{LirMa} that the $\Gamma^i{}_{jk}$ are constant on the fuzzy sphere, given that the $g_{ij}$ have to be.  To do some physics we
focus on the static rotationally invariant case where $g_{ij} = k\delta_{ij}$,  for a positive constant $k$. In this case it follows from the above that  ${\rm Ricci}= 0$ if and only if 
\[H(r)={1\over \beta(r)},\quad \beta(r) = \frac{1}{2}({1\over k}+1) - {3\over4}k + \frac{c_1}{r^2},\]
where $c_1$ is an arbitrary constants. The values 
\begin{equation}\label{kval}  k = {1\over 3}(\sqrt{7}-1 ),\quad c_1=- r_H^2\end{equation}
give the form of $\beta$ for the Tangherlini black hole metric of mass $M$, namely
\begin{equation}\label{tangrh}\beta(r) = 1 - \frac{r_H^2}{r^2},\quad r^2_H = {8\over 3 }G_5M,\end{equation}
but note that the latter only makes sense in 5D spacetime due to an extra length dimension in the Newton constant $G_5$. We are thinking of our model as 4D so we will {\em not} take this value but just work with $r_H$ as a free parameter. A different value of $k$ can be absorbed in different normalisation of the $t,r$ variables while $r_H$ is more physical.
 
 The quantum geometric structures in this `fuzzy black hole'  Ricci flat case are
\begin{align} g&= -(1-{r_H^2\over r^2})\extd t \tens \extd t + (1-{r_H^2\over r^2})^{-1} \extd r \tens \extd r + r^2 k s^i\tens s^i,\\
(\extd t,\extd t) &=-{r^2\over r^2 - r_H^2} ,\quad  (\extd r,\extd r) = 1 - \frac{r_H^2}{r^2}, \quad (s^i,s^j) = {\delta^{ij} \over kr^2},\\
\nabla \extd t &=- {r_H^2\over r(r^2 - r_H^2 )} \extd r \tens_s \extd t, \\
 \nabla \extd r &= {r_H^2\over r(r^2 - r_H^2)} \extd r \tens \extd r -{r_H^2\over r^3}\left(1-{r_H^2\over r^2}\right)\extd t \tens \extd t+ rk\left(1 - {r_H^2\over r^2}\right)  s^i \tens s^i, \\
 \nabla s^i &= -\frac{1}{2}\epsilon^{i}{}_{jk} s^j\tens s^k - \frac{1}{r} \extd r \tens_s s^i, \\
 {\rm R}_\nabla\extd t & = -{3r_H^2\over r^2(r^2-r_H^2)} \extd t \wedge \extd r \tens \extd r + \left( {r_H\over r} \right)^2k\extd t \wedge s^i \tens s^i,\\
 {\rm R}_\nabla\extd r & = \left( {r_H\over r} \right)^2 k\extd r \wedge s^i \tens s^i  + 3r_H^2{r^2-r_H^2\over r^6} \extd r \wedge \extd t \tens \extd t, \\
 {\rm R}_\nabla s^i & =  \left(-{1 \over 4} + k\left( 1 - {r_H^2\over r^2} \right)  \right) s^i \wedge s^j \tens s^j + \left( {r_H\over r} \right)^2{1\over r^2 - r_H^2} s^i \wedge \extd r \tens \extd r\nonumber\\&\quad + {r_H^2\over r^6}(r_H^2-r^2)s^i \wedge \extd t \tens \extd t,\\ \label{eq:lap_fbh}
 \Delta &= - \left(1 - {r_H^2\over r^2}\right)^{-1}\del_t^2 +  \left( {3\over r} - {r_H^2\over r^3} \right)\del_r + \left(  1 - {r_H^2\over r^2} \right)\del_r^2 + {1\over kr^2}\sum_i\del^2_i.\end{align}

For comparison,  the classical Tangherilini 5D black hole metric has the form 
\[  g = - \left(1 - {r_H^2\over r^2}\right) \extd t \tens \extd t + \left(1 - {r_H^2\over r^2}\right)^{-1}\extd r \tens \extd r  + r^2 g_{S^3} \] with the Laplacian 
\[  \Delta  = - \left(1 - {r_H^2\over r^2}\right)^{-1}\del_t^2 +  \left( {3\over r} - {r_H^2\over r^3} \right)\del_r + \left(  1 - {r_H^2\over r^2} \right)\del_r^2 + {1\over r^2}\Delta_{S^3}\]
where $g_{S^3}$ denotes the metric element on a unit $S^3$. We see that this has just the same form as our metric and Laplacian except that our unit fuzzy sphere Laplacian $\sum_i \del_i^2$ is replaced by the unit $S^3$ Laplacian
\[ \Delta_{S^3} = {1\over \sin^2{\psi}}\del_\psi(\sin^2{\psi}\del_\psi) + {1\over \sin^2{\psi}\sin{\theta}}\del_\theta(\sin{\theta}\del_\theta) + {1\over \sin^2{\psi}\sin^2{\theta}}\del^2_\phi \]
in standard angular coordinates. 

We will also be interested in the spatial geometry of the fuzzy black hole as a time slice with respect to the $t$ coordinate. This is easily achieved in our formalism. 

\begin{proposition}\label{fuzBHspatial} Defining the spatial geometry of the fuzzy black hole as a slice of the 4D geometry by setting $\extd t=0$, gives 
\begin{align*} g&=\beta^{-1}\extd r\tens\extd r+ k r^2 s^i\tens s^i,\\
\nabla\extd r&={r_H^2\over r^3\beta}\extd r\tens\extd r+ k r\beta s^i\tens s^i,\quad\nabla s^i=-{1\over 2}\eps^i{}_{jk}s^j\tens s^k-{1\over r}\extd r\tens_s s^i,\\
R_\nabla \extd r&=k({r_H\over r})^2\extd r\wedge s^i\tens s^i,\quad R_\nabla s^i=(k\beta-{1\over 4})s^i\wedge s^j\tens s^j+ {r_H^2\over  r^4\beta}s^i\wedge\extd r\tens\extd r,\\
{\rm Ricci}&= {3 r_H^2\over 2 r^4\beta}\extd r\tens\extd r+\left(k(1-{r_H^2\over 2r^2})-{1\over 4}\right)s^i\tens s^i,\quad S={3\over r^2}(1-{1\over 4k})
\end{align*} using the antisymmetric lift as usual and $\beta=1-{r_H^2\over r^2}$. The spatial Einstein tensor ${\rm Eins}={\rm Ricci}-{S\over 2}g$ comes out as
\[ {\rm Eins}={3\over 2 r^2\beta}({1\over 4 k}-\beta)\extd r\tens\extd r+{1\over 2}({1\over 4}-k(1+{r_H^2\over r^2}))s^i\tens s^i\]
and is conserved in the sense $\nabla\cdot{\rm Eins}=0$. \end{proposition}
 \proof  That setting $\extd t=0$ gives a QLC for the reduced metric and its curvature follows on general grounds but can be checked explicitly. The computation of Ricci is a trace of $R_\nabla$ as usual: we apply this to the second factors of $g$ and then apply $(\extd r,\extd r)=\beta^{-1}$, $(s^i,s^j)={\delta_{ij}\over k r^2}$ (and other cases zero) to the first two tensor factors. The Ricci scalar $S$ and Einstein tensor then follow. For its divergence, we first compute $\nabla{\rm Eins}$ by acting with $\nabla$ on each tensor factor but keeping its left-most output to the far left,
 \begin{align*} \nabla{\rm Eins}&={1\over 2}({1\over 4}-k(1+{r_H^2\over r^2}))(-{1\over r}s^i\tens s^i\tens\extd r)+\extd({3\over 2 r^2\beta}({1\over 4 k}-\beta))\tens\extd r\tens\extd r\\
 &\quad +{3\over 2 r^2\beta}({1\over 4 k}-\beta)({2 r_H^2\over r^3\beta}\extd r\tens\extd r\tens\extd r+ k r \beta s^i\tens s^i\tens\extd r)+\cdots\end{align*}
where $\cdots$ refers to terms that involve $\extd r\tens s^i$ or $s^i\tens \extd r$ in the first two tensor factors. The terms in $\nabla(s^i\tens s^i)$  with $s$'s in all tensor factors cancel. We then define $\nabla\cdot{\rm Eins}$ by applying $(\ ,\ )$ to the first two tensor factors to give
 \[ \nabla\cdot{\rm Eins}=\left(-{3\over 2 r^3 k}({1\over 4}-k(1+{r_H^2\over r^2}))+\beta({1\over 4}-k(1+{r_H^2\over r^2}))'+{3\over 2 r^2}({1\over 4k}-\beta){2r_H^2\over r^3\beta}+ {9\over 2 r^3}({1\over 4k}-\beta)\right)\extd r\]
 from the displayed terms taken in order. We then check that the function in brackets vanishes. \endproof

\subsection{Motion in the fuzzy black hole background}\label{secqmbh} In terms of physical implications,  since the radial form for the fuzzy black hole is  the same as that of the Tangherilini solution, we can apply the usual logic that $g_{00}=-(1+2\Phi)$ to first approximation contains the gravitational potential $\Phi$  per unit mass governing geodesic motion for a mass $m$ in the weak field limit. Therefore in our case, this should be 
\begin{equation}\label{fuzPhi}\Phi=-{r_H^2\over 2 r^2}\end{equation}
but, because we are thinking of this as a 4D model, we do not set $r_H$ to be the same as a   Tangherilini 5D black hole. Rather, we think of $r_H$ as the physical parameter and equate it for purpose of comparison with $r_H=2GM$ so that the horizon occurs at the same $r$ as for a Schwarzschild black hole of mass $M$. The weak field force law is no longer Newtonian gravity, having an inverse cubic form in $r$ according to $\Phi=-2 G^2 M^2/r^2$. This is rather different from modified gravity schemes such as MOND for the modelling of dark matter\cite{Mil} but could still be of interest. 

To properly justify the above, we should study geodesics, which is possible but not easy on quantum spacetimes. Here, we instead reach the same conclusion from the point of view of quantum mechanics as the non-relativistic limit of the Klein-Gordon equation
\[ \Delta \phi=m^2\phi.\]
As proof of concept, we first do this for a Schwarzschild black hole where $\beta(r)=1-{r_H\over r}$. The Laplacian is
\[ \Delta_{Sch}=-{1\over\beta}{\del^2\over\del t^2}+\Delta_r+ {1\over r^2}\Delta_{S^2};\quad \Delta_r:={1\over r^2}{\del\over\del r}(\beta r^2{\del\over\del r})\]
and $\Delta_{S^2}$ is normalised to have eigenvalues $\lambda_l=-l(l+1)$ on the spherical harmonics of degree $l\in \N\cup\{0\}$ for the orbital angular momentum. We focus on waves of fixed $l$ and look for solutions of the form
\[ \phi=e^{-\imath m t}\psi_l(t,r)\]
with $\psi_l$ slowly varying in $t$. Accordingly neglecting its double time derivative, the Klein-Gordon equation becomes the Schroedinger-like equation
\begin{equation}\label{bhsch} \imath\dot\psi_l=-{\beta\over 2m}\left(\Delta_r+{\lambda_l\over r^2}\right)\psi_l+ (\beta-1){m\over 2}\psi_l,\end{equation}
where $\beta(\Delta_r+{\lambda_l\over r^2})$ is a modification by $\beta$ of the $\R^3$ Laplacian on $\psi_l$ in polars, which we think of as the square of a modified momentum (the difference is anyhow suppressed at large $r$), and $(\beta-1)m/2=-GMm/r$ is the expected Newtonian potential for Schroedinger's equation in the presence of a point source of mass $M$.

Note that $e^{-\imath m t}$ is not itself a solution of the Klein-Gordon equation. Repeating the above but with reference to an actual solution would be analogous to finding the forces experienced by a particle in geodesic motion, where one only sees tidal forces. Continuing in the Schwarzschild case, we first solve (numerically) for spherical $l=0$ solutions of the form
\begin{equation}\label{bhphim} \phi=e^{-\imath \omega t}\phi_\omega(r);\quad ({\omega^2\over\beta}-m^2)\phi_\omega+\Delta_r\phi_\omega=0\end{equation}
with initial conditions specified at large $r$. We then look for a Schroedinger-like equation relative to   $\phi_\omega$ by solving the Klein-Gordon equation for solutions of the form 
\begin{equation}\label{compsi} \phi=e^{-\imath \omega t}\phi_\omega(r)\psi_l(t,r)\end{equation}
with $\psi_l$ of orbital angular momentum labelled by $l$ and slowly varying in $t$. This time we obtain
\begin{equation}\label{bhschcom} \imath\dot\psi_l=-{\beta\over 2\omega}(\Delta_r+{\lambda_l\over r^2}+2\beta{\phi'_\omega\over \phi_\omega}{\del\over\del r})\psi_l\end{equation}
with a new velocity-dependent correction but without the gravitational point source potential, as expected. 

The natural choice for reference field here to focus on the case $\omega=m$.  For large $r$, we can neglect $\beta'$ relative to $2/r$ in $\Delta_r$ and in this case one has an exact solution for $\phi_m$ in terms of Bessel I, K functions: we choose conditions which match to Bessel I, say, at large $r$. We assume $m>1/r_H$ so that the Compton wavelength is less than $r_H$. Then  $\phi_m'/\phi_m$ is barely oscillatory for larger $m$ and decays gradually as $r\to\infty$ according to
\begin{equation}\label{phimdif} {\phi'_m\over \phi_m}\approx \imath m \sqrt{r_H\over r-r_H},\quad r>> r_H.\end{equation}
\begin{figure}\[ \includegraphics[scale=0.5]{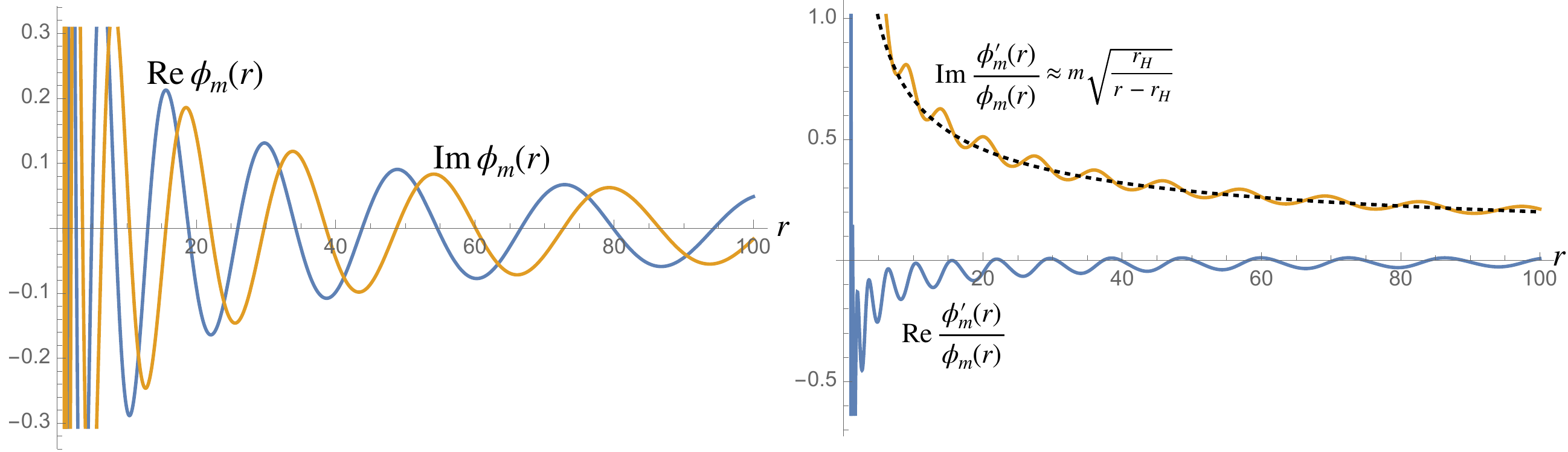}\] \caption{Radial solution $\phi_m(r)$ for the Klein-Gordon equation around a black hole shown for  $m=2/r_H, r_H=1$ and asymptotic form of $\phi_m'/\phi_m$ shown dashed\label{figBH}}\end{figure}
\begin{figure}\[ \includegraphics[scale=0.75]{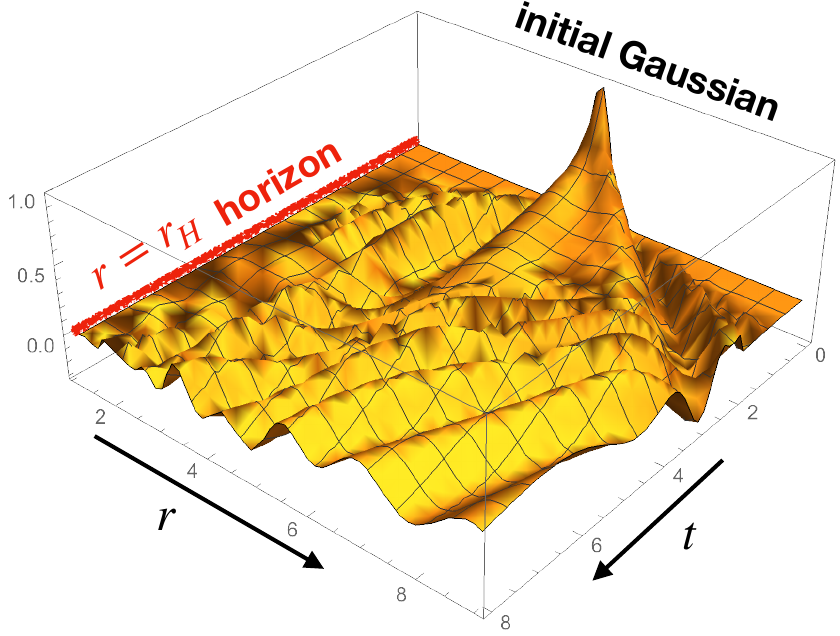}\] \caption{Schroedinger-like evolution relative to the $\phi_m$ in Figure~\ref{figBH}. We see an initial Gaussian centred at $r=5r_H$ evolving much as in quantum mechanics but decaying over time, with the essentially zero initial values at $r=1.1r_H,10r_H$ held fixed. \label{figBHsch}}\end{figure}
\noindent The actual numerical solution as illustrated in Figure~\ref{figBH} is similar, although more oscilliatory. We see that in this `comoving frame' from a Klein-Gordon point of view, we do not experience the main force of gravity but we do see a novel radial velocity term in the effective Schroedinger-like equation approximated as
\begin{equation}\label{bhcompsi} \imath\dot\psi_l\approx-{\beta\over 2m}(\Delta_r+{\lambda_l\over r^2})\psi_l-\imath\beta^{3\over 2}\sqrt{2GM\over r}\psi_l'\end{equation}
far from the horizon. Nearer the horizon, one needs to use the actual $\phi_m'/\phi_m$ to avoid an instability coming in from the horizon. A numerical solution for $\psi_l$ at $l=0$ using the actual values is  in Figure~\ref{figBHsch}, showing an initial Gaussian centred at $r=5r_H$ evolving much as in regular quantum mechanics but, unlike the latter, decaying over time. Some of the noise in the picture comes from the numerical approximation.

The above is for a regular black hole, but  one can make a similar analysis for the different radial equations for our fuzzy black hole and thereby justify (\ref{fuzPhi}), provided we know something about the operators $\sum_i\del_i^2$. 

\begin{proposition} ${1\over 2}\sum_i\del_i^2$ on the fuzzy sphere  has eigenvalues $\lambda_l=-l(l+1)$ as for the classical $\Delta_{S^2}$, with eigenspaces 
	\[ H_l=\{ x^{i_1}x^{i_2}\cdots x^{i_l}f_{i_1\cdots i_l}\ |\ f\ {\rm totally\ symmetric\ and\ traceless}\}.\]
\end{proposition}
\proof Here, as vector spaces, $\C[x^1,\cdots,x^n]=\C[su_2^*]\cong U(su_2)$ by the Duflo map (as for any Lie algebra). This sends a commutative monomial in the $x^i$ to  an average of all orderings of its factors (it is an isomorphism because, although there are nontrivial commutation relations in the enveloping algebra, these are strong enough to reorder at the expense of lower degree.) This map is covariant for the coadjoint and adjoint actions, in our case, of $su_2$, and therefore descends to an isomorphism between polynomial functions on the classical sphere in cartesian coordinates on one side, and the fuzzy sphere $\C_\lambda[S^2]$ on the other side. Moreover, $\del_k x_i=\eps_{ijk}x^j$ for our differential calculus on the latter acts as orbital angular momentum. Hence $\sum_i\del_i^2$ acts as the quadratic Casimir and can be computed on the classical sphere,  where it decomposes the polynomial functions into the spherical harmonics of each degree $l$. These then correspond to the $H_l$ as stated. One can check this directly on the fuzzy sphere on low degrees by hand, to fix the normalisation. For example, on degree $l=1$, we have $\sum_k\del_k^2x^i=\del_k\eps_{ijk}x^j=\eps_{jmk}\eps_{ijk}x^m=-2x^i$.  \endproof

Thus, we can solve the Laplacian and look at the non-relativistic limits by the same methods as we illustrated for the Schwarzschild black hole. The only difference is that  the functions have values $\psi_l(t,r) \in \C_\lambda[S^2]$, but the differential equations themselves in $t,r$ are purely classical according to
\[ \Delta_{fuz}=-{1\over\beta}{\del^2\over\del t^2}+\Delta_r+ {2\over k r^2}\lambda_l;\quad \Delta_r:={1\over r^3}{\del\over\del r}(\beta r^3{\del\over\del r})\]
with $\beta=1-r_H^2/r^2$. Taking $e^{-\imath m t}$ as reference gives the same form as (\ref{bhsch}) but with $2\lambda_l\over k r^2$ in a modified effective spatial Laplacian. Then  $(\beta-1)m/2=-2G^2M^2 m/r^2$ for the gravitational potential energy in agreement with (\ref{fuzPhi}). 

Next, for the `comoving' version, the $l=0$ solutions of the Klein-Gordon equation are given by  solving (\ref{bhphim}) as before and relative to this,  slowly-varying $\psi_l$ defined by (\ref{compsi}) obey the Schroedinger-like equation (\ref{bhschcom}) but now with $2\lambda_l\over k r^2$ in place of $\lambda_l\over r^2$. Focussing on the $\omega=m$ case, the main difference now is that $\phi_m$ decays more rapidly and in first approximation, if we leave out the $\beta'$ term in $\Delta_r$, is now solved by 
\[ \phi_m(r)\propto \frac{\left(r^2-r_H ^2\right)^{\frac{1}{2}\pm\frac{1}{2} \sqrt{1-m^2 r_H ^2}}}{ r^2}.\]
We focus on the $+$ case of the square root, which  leads for $m>>1/r_H$ to a fair approximation
\begin{figure}\[ \includegraphics[scale=0.76]{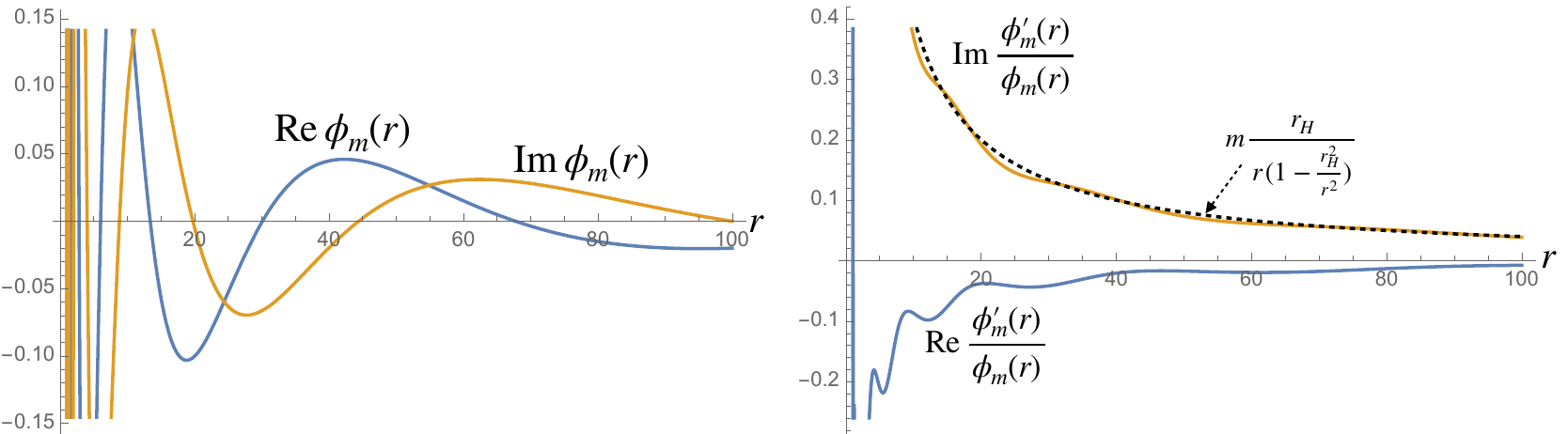}\] \caption{Radial solution $\phi_m(r)$ for the Klein-Gordon equation around a fuzzy black hole shown for  $m=4/r_H, r_H=1$, and function $\phi'_m(r)/\phi_m(r)$. \label{figBHF}}\end{figure}
\[ {\phi_m'\over\phi_m}\approx  \imath m {r_H\over r (1-{r_H^2\over r^2})},\quad  r>> r_H\]
as illustrated in Figure~\ref{figBHF}. As a result, the long range Schroedinger-like equation is
\[\imath\psi \approx - {\beta\over 2m}\left( \Delta_r + {2\lambda_l\over kr^2} \right)\psi_l - \imath\beta{2GM\over r}\psi_l'\]
if we use the Schwarzschild value of $r_H$, showing a coupling to the velocity term of the same size as the usual gravitational potential per unit mass.  As before, near the horizon we need the actual $\phi_m'/\phi_m$ values for stability of the solutions. An initial Gaussian breaks up and decays over time, looking much as before. 

Finally, although we have used the Schwarzschild value of $r_H$ for purposes of comparison,  since the geometry is asymptotically flat, we could naively try to define an actual ADM mass by copying its physical formulation in terms of the Einstein tensor of the spatial geometry\cite{Ash,Cru,Mia}, which in spherical polars amounts to the limit $r\to\infty$ of 
\begin{align*} M(r)&={2\over G(n-1)(n-2)\Omega_{n-1}}\int_{S^{n-1}_r}{\rm Eins}(r\del_r, \sqrt{\beta}\del_r)\extd^{n-1}\Omega\\
&={2\over G(n-1)(n-2)}r^{n-1}{\rm Eins}(r\del_r, \sqrt{\beta}\del_r)\end{align*}
for a spatial geometry of dimension $n$. Here there is a factor $-2$ compared to the usual definition because our Ricci and hence Einstein tensor conventions reduce in the classical case to $-{1\over 2}$ of the usual ones. $\Omega_{n-1}$ is the volume of a unit sphere of dimension $n-1$ and we integrate with measure $\extd^{n-1}\Omega$ over the sphere $S^{n-1}_r$ at radius $r$. The conformal Killing vector field in the general formula in \cite{Mia} is just $r\del_r$ in our case and the unit outward normal vector field is $\sqrt{\beta}\del_r$ given the form of the spatial metric. As everything is rotationally invariant, the integration merely gives a factor $r^{n-1}\Omega_{n-1}$.  For a usual Schwarzschild black hole of mass $M$, the  Einstein tensor of the spatial geometry in our conventions can be extracted from  \cite[Cor.~9.9]{BegMa} to find
\[ {\rm Eins}_{Sch}(r\del_r,\sqrt{\beta}\del_r)={1\over 2 r \sqrt{\beta}}(1-\beta),\quad M(r)={r_H\over   2G\sqrt{\beta}}\to {r_H\over 2G}\]
as expected for the Schwarzschild $\beta(r)=1-{r_H\over r}$. If we now use the fuzzy quantum black hole spatial geometry in Proposition~\ref{fuzBHspatial}, the radial sector is completely classical so it makes sense to read off ${\rm Eins}(\del_r,\del_r)$ as the coefficient of $\extd r\tens\extd r$, resulting in our case in 
\[ {\rm Eins}_{fuz}(r\del_r,\sqrt{\beta}\del_r)={3\over 2 r \sqrt{\beta}}({1\over 4k}-\beta),\]
which, since $\beta(r)=1-{r_H^2\over r^2}$ and $k\ne {1\over 4}$, results in $M(r)\to \infty$. If we took $k={1\over 4}$ then we would not have a Ricci flat metric in the spacetime quantum geometry and we would get $M(r)\to 0$, which is not reasonable either. These problems are a consequence of the dimension jump in the quantum model, evidently requiring a more sophisticated approach to ADM mass. Indeed, if we were to set $n=4$ and $k={1\over 4}$ then we would obtain $M(r)\to {r_H^2\over 2G}$, which is rather close to the value (\ref{tangrh}) for a classical 5D black hole. 

\section{Black hole  with the discrete circle}\label{secBHZn}

We now consider the same ideas as in the preceding section but for a 3D spacetime metric with $S^1$ in polar coordinate replaced by the discrete group $\Z_n$. The 2D FLRW model with $S^1$ replaced by $\Z_n$ was done in \cite{ArgMa} and we use the same notations. Briefly, $i\in \Z_n$ now labels the vertices of a polygon as an integer mod $n$. The `step up' and `step down' partial derivatives as $\del_\pm=R_\pm-\id$, where $(R_\pm f)(i)=f(i\pm 1)$ and $e^\pm$ are the associated invariant 1-forms with $e^+{}^*=-e^-$. The calculus on $\Z_n$ is not commutative as $e^\pm f= R_\pm(f)e^\pm$ for a function $f$, the $e^\pm$ anticommute with each other and the exterior derivative is $\extd f=(\del_\pm f)e^\pm$ (sum over $\pm$) and $\extd e^\pm=0$. The classical limit $n\to \infty$ can be seen as a circle with a noncommutative 2D calculus which is the classical calculus on $S^1$ extended by a 1-form $\Theta_0$. The latter has no classical analogue but can be viewed as normal to the circle when embedded in a plane\cite{ArgMa}, but without an actual normal variable. The natural invariant metric on the polygon is $-e^+\tens_s e^-$.

Now the spacetime coordinate algebra is  $A=C^\infty(\R\times\R_{>0})\tens \C(\Z_n)$ with $t,r$ for the time and radial classical variables, and we consider a static Schwarzschild-like metric of the form
\begin{equation} \label{eq:bhm-pol} g = -\beta(r)\extd t \tens \extd t + H(r) \extd r \tens \extd r - \alpha_{ab}(r,i) e^a\tens e^b.\end{equation}
Invertibility of the metric requires centrality, which dictates $\alpha_{ab}(r,i) = \alpha_{a}(r,i)\delta_{ab^{-1}}$ for some real-valued functions $\alpha_{a}$. We also require edge-symmetry $\alpha_a = R_a(\alpha_{a^{-1}})$ so that the length of in each edge $\bullet_i-\bullet_{i+1}$ for the $\Z_n$ at radius $r$ is the same in either direction, namely given by some real function $a(r,i)$ according to 
\begin{equation}\label{alphaa} \alpha_+(r,i)=a(r,i),\quad \alpha_-(r,i)=R_-a(r,i).\end{equation} 
We limit attention to this form of metric.     

 We take analogous conditions on the tensor product calculus as in the previous section, in the sense  that the functions of the time $t$, radius $r$ as well as $\extd t,\extd r$ are classical and graded-commute with everything. In view of this, and in line with  \cite{ArgMa} and with the fuzzy case above, {\em we make the simplifying assumption that the connection braiding $\sigma$ among the differentials $\extd r,\extd t$ and between them and $e^\pm$ is just the flip map}. In this case, the most general form of a potential bimodule connection turns out to be
 \begin{align*}
	\label{eq:general SCF}
	 \nabla e^a &= -\Gamma^a{}_{bc}e^b\tens e^c  + \nu^a{}_b \extd t \tens_s e^b + \gamma^a{}_b \extd r \tens_s e^b, 
\\
\nabla \extd t &= \xi_{ab}e^a\tens e^b + b \extd t \tens \extd t + c \extd r \tens \extd r+ h \extd r \tens_s \extd t, 
\\
\nabla \extd r&= A_{ab}e^a\tens e^b + B \extd t \tens \extd t + C \extd r \tens \extd r+ D \extd r \tens_s \extd t, 
\end{align*}
 where the coefficients are elements of the algebra $A$ and of the form
\begin{equation}
	\nu^a{}_b = \nu_a\delta_{a,b^{-1}}, \quad \gamma^a{}_b = \gamma_a\delta_{a,b^{-1}}, \quad A_{ab}  = A_{a}\delta_{a,b^{-1}},\quad \xi_{ab} = \xi_{a}\delta_{a,b^{-1}}.
\end{equation}

We now analyse when such a bimodule connection is a QLC. The requirement to be torsion free comes down to
\begin{equation}
\label{eq:tor-free-pol}
	A_{ab} = A_{ba}, \quad \xi_{ab} = \xi_{ba}, \quad \Gamma^a{}_{bc} = \Gamma^a{}_{cb}, \quad \wedge (\id+\sigma)(e^a\tens e^b) = 0,
\end{equation}
while to be metric compatible comes down to the 13 equations:
\begin{align*}
	\extd r \tens \extd t \tens \extd t &: \del_r\beta + 2\beta h = 0, \\
	\extd r^{\tens 3} &: \del_rH + 2H C = 0, \\
	\extd t^{\tens 3} &: 2\beta b = 0, \\
	\extd r \tens \extd t \tens \extd r/\extd r \tens \extd r \tens \extd t &: -\beta c + HD = 0, \\
	\extd t \tens \extd t \tens \extd r/\extd t \tens \extd r \tens \extd t &: -\beta h + HB = 0, \\
	\extd t \tens \extd r \tens \extd r &: 2HD = 0, \\
	\extd r \tens e^a \tens e^b & : -\del_r \alpha_{ab} - \alpha_{cb}\gamma^c_a - \alpha_{ac}R_{a}(\gamma^c_b) =0, \\
	e^a \tens e^b \tens \extd t & : -\beta \xi_{ab} - \alpha_{cd}R_c(\nu^{d}{}_{f}) \sigma^{cf}{}_{ab} = 0, \\	
	e^a \tens \extd t \tens e^b &: -\beta \xi_{ab} - \alpha_{cb}\nu^{c}{}_{a} = 0, \\
	e^a \tens e^b \tens \extd r & : HA_{ab} - \alpha_{cd} R_c(\gamma^{d}{}_{f}) \sigma^{cf}{}_{ab}= 0, \\	
	e^a \tens \extd r \tens e^b  & : HA_{ab} - \alpha_{cb} \gamma^{c}{}_{a}  = 0, \\	
	e^a \tens e^b \tens e^c & : -\del_a\alpha_{bc}  -\alpha_{dc}\Gamma^{d}{}_{ab} -\alpha_{df} R_d(\Gamma^{f}{}_{gc})\sigma^{dg}{}_{ab} = 0, \\	
	\extd t \tens e^a \tens e^b & : -\alpha_{cb}\nu^{c}{}_{a} - \alpha_{ac}R_a(\nu^{c}{}_{b}) = 0. 	
\end{align*}
The 1st and 2nd equations gives $h,C$ respectively, and these  together with the 5th equation give  $B$, as 
\begin{equation}
	h = - {1\over 2\beta }\del_r \beta, \quad C = -{1\over 2H}\del_r H, \quad B = -{1\over 2H}\del_r\beta. 
\end{equation}
The 3rd, 6th and 4th equations imply that $c = b = D =  0$.  Next, the  9th and 11th equations tell us that
 \begin{equation} \label{eq:gam_nu}
 	\nu_a = -{\beta \xi_a\over\alpha_a}, \quad  \gamma_a = {HA_a\over \alpha_a}, 
 \end{equation}
 while, given the edge-symmetry, the 13th and 7th equations reduce to
 \begin{equation} \label{eq:R_gam_nu}
    \nu_a + R_a(\nu_{a^{-1}})=0, \quad \gamma_a + R_a(\gamma_{a^{-1}}) = -{\del_r\alpha_a \over \alpha_a }.
 \end{equation}

Given that the 12th equation for metric compatibility and the torsion-freeness condition are the same as for the polygon in \cite{ArgMa}, we are led to take $\Gamma^{a}{}_{bc}$ at each radius $r$ the same as for the QLC $\nabla^{\Z_n}$ on the polygon found there. This has
\[ \nabla^{\Z_n}e^+=(1-\rho)e^+\tens e^+,\quad \nabla^{\Z_n}e^-=(1-R_-^2\rho^{-1})e^-\tens e^-,\quad \rho(r,i)={a(r,i+1)\over a(r,i)}\]
and its braiding obeys $\sigma(e^\pm\tens e^\mp)=e^\mp\tens e^\pm$, in which case the 8th and 10th metric compatibility equations become
\begin{equation} \label{eq:a-A}
	 R_{a}(\nu_{a^{-1}}) = -{\beta\over \alpha_{a}} \xi_{a^{-1}}, \quad R_{a}(\gamma_{a^{-1}}) = {HA_{a^{-1}}\over \alpha_{a}}.
\end{equation}
Using the first  of (\ref{eq:gam_nu}) and (\ref{eq:a-A}) in (\ref{eq:R_gam_nu}) leads us to  $\xi_a = -\xi_{a^{-1}}$, which together with the second half of the torsion-freeness conditions (\ref{eq:tor-free-pol}) requires $\xi_a = 0$, and as consequence $\nu_a=0$. Similarly, inserting the second half of (\ref{eq:gam_nu}) and (\ref{eq:a-A}) in (\ref{eq:R_gam_nu}) produces 
\begin{equation} \label{eq:A_cond}-A_a - A_{a^{-1}} =  {\del_r\alpha_a\over H}.\end{equation}
In summary, for a QLC, it only remains to solve for $A_a,\gamma_a$ subject to such residual equations, with the other coefficients zero or determined. It also remains to impose reality in the form of $\nabla$ $*$-preserving.

\begin{proposition} Assuming a static edge-symmetric central metric (\ref{eq:bhm-pol}) and $\sigma$ the flip on generators involving $\extd r,\extd t$ leads to a $*$-preserving QLC if and only if $\del_-\del_r \alpha_a=0$ (which needs the underlying $a(r,i)$ to be the sum of a function of $r$ and a function of $i$). The $*$-preserving QLC with real coefficients is then unique  and given by
\begin{align*}
	\nabla \extd t =& -\frac{1}{2\beta} \del_r\beta \extd r \tens_s \extd t \\
	\nabla \extd r =& -\frac{1}{2H} \del_rH \extd r \tens \extd r - {\del_r\alpha_+\over 2H} e^+ \tens e^- - {\del_r\alpha_-\over 2H} e^- \tens e^+ - \frac{1}{2H} \del_r\beta \extd t \tens \extd t\\
	\nabla e^{\pm} =&   \nabla^{\Z_{n}}e^\pm- \frac{1}{r} \extd r \tens_s e^{\pm}  .
\end{align*}
\end{proposition}
\proof The $*$-preserving conditions for $\nabla$ include conditions on $\Gamma$ which coincide at each $r$ with those for a QLC on $\Z_n$ as in \cite{ArgMa}, for which the solution is unique, so we are forced to this choice for $\Gamma$. The remaining $*$-preserving conditions require $B,C,h$ to be real-valued, which already holds  because they are functions of the metric coefficients, together with the conditions
\begin{align} \label{eq:preserv_pol_a}
	\sum_a ( \overline{A}_{a^{-1}}\sigma(e^{a^{-1}}\tens e^{a}) - A_a e^a\tens e^{a^{-1}}) = 0, \quad \overline{\gamma}_a = R_a(\gamma_{a^{-1}}), \\
	\label{eq:preserv_pol_b}
	\sum_a ( \overline{\xi}_{a^{-1}}\sigma(e^{a^{-1}}\tens e^{a}) - \xi_a e^a\tens e^{a^{-1}}) = 0, \quad \overline{\nu}_a = R_a(\nu_{a^{-1}}).
\end{align}
 The conditions (\ref{eq:preserv_pol_b}) are trivially fulfilled, while  the second half of (\ref{eq:preserv_pol_a}) implies $\overline{A}_a = A_{a^{-1}}$, which together with the form of the braiding map $\sigma$ solves the first half of (\ref{eq:preserv_pol_a}).  In this case,  (\ref{eq:A_cond}) takes the form
\begin{equation}  -A_a - \overline{A}_{a} = {\del_r{\alpha_a}\over H}.\end{equation} 
The second halves of (\ref{eq:gam_nu}) and (\ref{eq:a-A}) together with the edge-symmetric condition, tell us that $A_a = R_{a^{-1}}(A_a)$ and hence that $A_a$ is independent of the discrete variable, i.e., just function of $r$. In this case, we must have 
\[ A_{\pm} = -{\del_r\alpha_{\pm}\over 2H} \pm \imath y(r), \quad \gamma_\pm = -{\del_r\alpha_\pm\over 2\alpha_\pm} \pm \imath {Hy(r)\over \alpha_\pm}\]
for some function real-valued function $y(r)$. It is natural at this point to set $y(r) = 0$ so as to keep coefficients real, and we do this now (this was also done at the parallel point in \cite{ArgMa}).  Another consequence of $A_\pm$ being constant in the polygon is $\del_\pm A_\pm = 0$, which lead us to $\del_\pm{\del_r\alpha_a} = 0$. This corresponds to restricting underlying metric function $a(r,i)$ in (\ref{alphaa}).  
\endproof
 
This is a general result, but we now restrict attention to the $\Z_n$-invariant metric where $a(r,i)$ is independent of $i$ and moreover of the expected radial form.

\begin{theorem}The static $\Z_n$-invariant Schwarzschild-like metric   
	\[ g = -\beta(r)\extd t \tens \extd t + H(r) \extd r \tens \extd r - r^2e^+\tens_s e^-\]
has a canonical $*$-preserving QLC,  
	\begin{align*}
		\nabla \extd t =& -\frac{1}{2\beta} \del_r\beta \extd r \tens_s \extd t, \\
		\nabla \extd r =& -\frac{1}{2H} \del_rH \extd r \tens \extd r - {r\over H} e^+ \tens_s e^- - \frac{1}{2H} \del_r\beta \extd t \tens \extd t,\\
		\nabla e^{\pm} =&   - \frac{1}{r} \extd r \tens_s e^{\pm}  
	\end{align*}
with the corresponding Ricci scalar and Laplacian 
\begin{align*}
S =& {1\over 2H\beta}\del^2_r\beta - {1\over 4H\beta^2}(\del_r\beta )^2 -{1\over 4H^2\beta}\del_rH\del_r\beta - {1\over rH^2}\del_rH + {1\over rH\beta}\del_r\beta + {1\over r^2H}, \\
\Delta =& {2\over r^2}(\del_+ + \del_-) - {1\over \beta}\del^2_t  + {1\over H }\del^2_r  +\left( {2\over rH} - {1\over 2 H^2}\del_rH + {1\over 2H\beta}\del_r\beta \right)\del_r.
\end{align*}	
This is Ricci flat if and only if
\begin{equation}
H(r) = {1\over \beta(r) }, \quad \beta(r) = {r_H \over r}
\end{equation}
for some constant $r_H$ of length dimension. 
\end{theorem}
\proof
Taking $\alpha_\pm=r^2$ in the preceding proposition immediately gives the canonical  QLC stated. Its associated curvature comes out as 
\begin{align*}
	R_\nabla\extd t &= {1\over 2\beta}\left( \del^2_r\beta - {1\over 2\beta}(\del_r\beta)^2 - {1\over 2H}\del_rH\del_r\beta \right)\extd t \wedge	\extd r \tens \extd r - {r\over 2H\beta}\del_r \beta \extd t \wedge e^+\tens_s e^-, \\
	R_\nabla e^{\pm} &= -{\del_r H\over 2rH} e^{\pm}\wedge \extd r \tens \extd r - {1\over 2rH}\del_r \beta e^{\pm}\wedge \extd t \tens \extd t - {1\over H} e^{\pm}\wedge e^{\mp}\tens e^{\pm}, \\
	R_\nabla \extd r &= {1\over 2H}\left( {1\over 2H} \del_r\beta \del_r H  - \del^2_r\beta + {1\over 2\beta}(\del_r\beta)^2 \right) \extd r \wedge \extd t \tens \extd t - {1\over2}r\del_r\beta \extd r \wedge e^{+}\tens_s e^-.
\end{align*}	
Taking the antisymmetric lift of products of basic 1-forms and tracing gives the associated Ricci tensor
\begin{align*}
	{\rm 2Ricci} &= \left( {1\over 2\beta}\del_r^2\beta - \left({\del_r\beta\over 2\beta}\right)^2 -{1\over 4H\beta}\del_rH\del_r\beta - {1\over rH}\del_rH \right)\extd r \tens \extd r \\
	&\left( - {r\over 2H\beta}\del_r\beta + {r\over 2H^2}\del_rH - {1\over H} \right) e^+\tens_se^- \\&+\left( {1\over 4H^2}\del_r\beta\del_rH - {1\over 2H}\del_r^2\beta + {1\over 4H\beta} (\del_r\beta)^2 - {\del_r\beta\over rH}\right)\extd t \tens \extd t
\end{align*}
The Ricci scalar and Laplacian follows on application of the inverse metric. We then solve for ${\rm Ricci}=0$. The calculations are straightforward and are omitted. 
\endproof 

 The quantum geometric structures in the `discrete black hole' Ricci-flat case are
\begin{align}\label{eq:metric_pol_rii_flat}
g =& -{r_H\over r}\extd t \tens \extd t + {r\over r_H} \extd r \tens \extd r - r^2e^+\tens_s e^-,\\ 
	(\extd t,\extd t) &=-{r_H\over r} ,\quad  (\extd r,\extd r) = {r\over r_H}, \quad (e^\pm,e^\mp) = -{1 \over r^2},\\
	\nabla \extd t =&  {1\over 2r} \extd r \tens_s \extd t,\\
	\nabla \extd r =& -{1\over 2r} \extd r \tens \extd r - r_H e^+ \tens_s e^- + {r^2_H\over 2r^3} \extd t \tens \extd t,\\
	\nabla e^{\pm} =&   - \frac{1}{r} \extd r \tens_s e^{\pm},  \\
	{\rm R}_\nabla \extd t =& {1\over r^2} \extd t \wedge \extd r \tens \extd r + {r_H\over 2r} \extd t \wedge e^+\tens_s e^-,\\
	{\rm R}_\nabla \extd r =& -{r_H^2\over r^4} \extd r \wedge \extd t \tens \extd t
	+{r_H\over 2r}\extd r \wedge e^+ \tens_s e^-, \\
	{\rm R}_\nabla e^{\pm} =& -{1\over 2 r^2 }e^{\pm} \wedge \extd r \tens \extd r + {r_H^2\over 2r^4}e^{\pm} \wedge \extd t \tens \extd t \mp {r_H\over r} e^+ \wedge e^- \tens e^\pm,\\
\label{disclap}	\Delta =&  -{r\over r_H}\del_t^2 + {r_H\over r}\del_r^2 + {r_H\over r^2}\del_r+{2\over r^2}(\del_+ + \del_-).
\end{align}
To keep the signature, we can take $r_H>0$ and we will analyse this case first. However, to approximately match the inside of a black hole, we should also analyse the case $r_H=-2GM<0$ with the physical roles of $r,t$ interchanged.

We also note that $\beta=H=1$ leads to 
\begin{align*} g&=-\extd t\tens\extd t+ \extd r\tens\extd r - r^2 e_+\tens_s e_-,\quad {\rm Ricci}=-{1\over 2}e^+\tens_s e^-,\quad S={1\over r^2},\\
 \Delta &=- \del^2_t  + \del^2_r  +{2\over r} \del_r  + {2\over r^2}(\del_+ + \del_-),\end{align*}
which is more like the spacetime Laplacian in 3 spatial dimensions, again showing the dimension jump and the
constant curvature at each fixed radius and time. Here $S^1$ behaves more like $S^2$ in polar coordinates, just with $2(\del_++\del_-)$ in the role of the angular Laplacian. 

\subsection{Klein-Gordon equation on the discrete-circle black hole for $\beta(r)>0$.}\label{kg_BhPoly}

Here, we analyse the case of the length scale $r_H>0$ in the Laplacian (\ref{disclap}) found for the discrete black hole above in `polar coordinates' form. The eigenvalues of the angular Laplacian $\del_++\del_-$ are labelled by $l\in \Z_n$ and given by 
\[ \lambda_l=q^l+q^{-l}-2=2(\cos({2\pi l\over n})-1)=-4\sin^2({\pi l\over n});\quad q=e^{2\pi\imath\over n}\]
with eigenfunctions $q^{il}$. If we followed the format of Section~\ref{secqmbh}, we might first consider the `quantum mechanical' solutions of Klein-Gordon equations  $\Delta \phi=m^2 \phi$ of the form
\[ \phi=e^{-\imath m t}\psi_l(t,r)\]
of orbital angular momentum $l$ and slowly varying in $t$. This is not particularly justified from the form of the metric but leads to 
\[ \imath\dot\psi=-{r_H\over 2m r} \left(\Delta_r+ {2\lambda_l\over r^2 }\right)\psi_l +  (1-{r_H\over r}){m\over 2}\psi_l;\quad \Delta_r= {r_H\over r^2}\del_r(r\del_r).\]
The mass term has not cancelled from the Klein-Gordon equation due to the $r_H/r$ factor in the $\extd t\tens\extd t$ term in the metric, except in the vicinity of $r\approx r_H$. 

\begin{figure}\[ \includegraphics[scale=0.78]{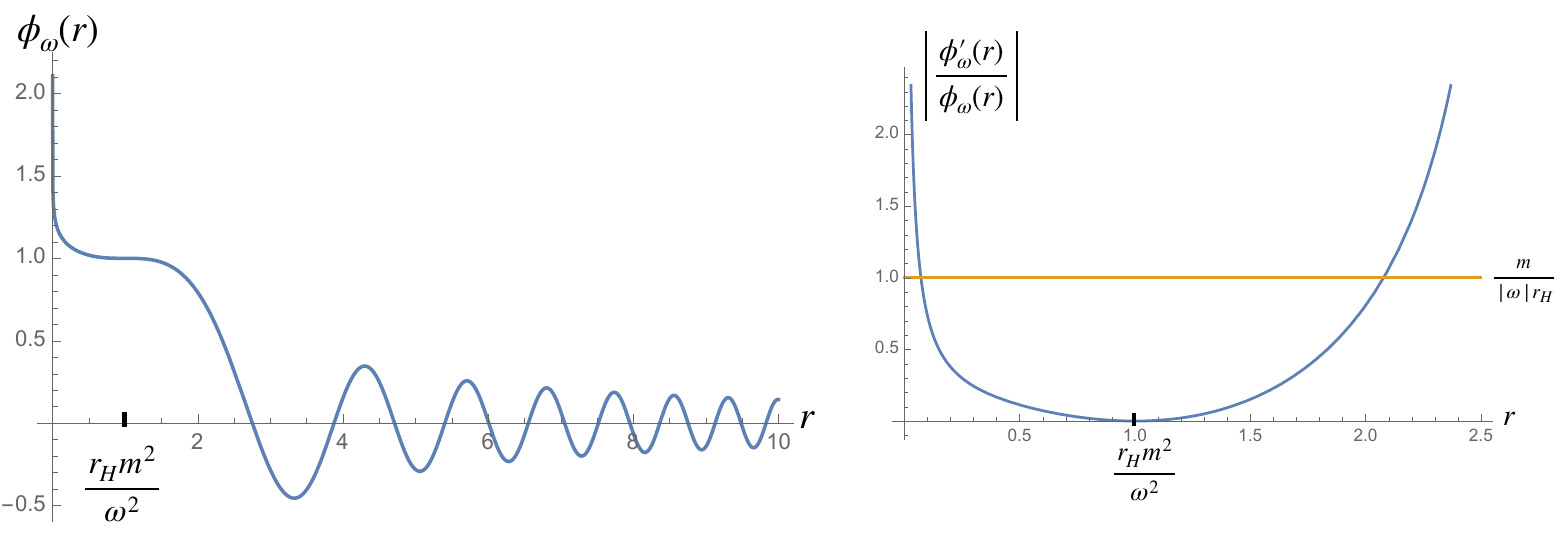}\] \caption{Solution of Klein-Gordon equation for $l=0$ and  $\omega=m=r_H=1$, with Cauchy boundary condition at $r_0=r_H{ m^2\over\omega^2}$.\label{kgdisc}}\end{figure}

Here it makes more sense to look in the `comoving' case where we start with an $l=0$ solution of the Klein-Gordon equation of the form 
\[ \phi=e^{-\imath\omega t}\phi_\omega;\quad  \phi_{\omega}''+ { 1\over r}\phi'_{\omega}+({r^2\over r_H^2}\omega^2-{r\over r_H}m^2)\phi_{\omega}=0.\]
A generic solution for $\omega=m=r_H=1$ is shown in Figure~\ref{kgdisc}, which illustrates that we can have an extended region where $\phi_\omega$ is approximately constant, here with  boundary condition  
\[ \phi'_\omega(r_0)=0,\quad \phi_\omega(r_0)=1;\quad r_0:=r_H {m^2\over\omega^2}.\]
This results in 
\[ \left|{\phi'_\omega(r)\over\phi_\omega(r)}\right|< {m \over |\omega|r_H},\quad r\approx  r_0\]
for a reasonable range around the central value, as illustrated in the second half of the figure. An obvious choice would be $\omega=m$ and hence $r_0=r_H$, but we can choose other $\omega$ to have other central values $r_0$.

Next, we use this as reference and look for solutions of the Klein-Gordon equations of the form $\phi=e^{-\imath \omega t}\phi_\omega(r)\psi_l(t,r)$ with $\psi_l$ in the $\lambda_l$ eigenspace and slowly varying in $t$. Discarding $\ddot\psi_l$ terms, we have
\[\imath \dot\psi_l=-{r_H\over 2\omega r}\left(\Delta_r+{2\phi'_\omega\over\phi_\omega} {r_H\over r}\del_r+ {2\lambda_l\over r^2}\right)\psi_l\]
and hence in any regime where the $\phi'_\omega/\phi_\omega$ term can be neglected,  we have approximately
\[ \imath\dot\psi_l \approx   -{r_H\over 2\omega r}\left(\Delta_r+ {2\lambda_l\over r^2}\right)\psi_l \]
as an effective Schroedinger-like equation. We still have an expected scale factor out front, but now the unwanted mass terms are absent, i.e. this looks more like free motion as expected.

We can go further and replace $r$ by  a new variable 
\[ \rho(r)={r^2\over 2 r_H},\quad {\del\over\del r}={r\over r_H}{\del\over\del \rho},\quad {\del^2\over\del r^2}={\del\over\del r}\left({r\over r_H}{\del\over\del \rho}\right)={r^2\over r_H^2}{\del^2\over\del \rho^2}+{1\over r_H}{\del\over\del\rho},\]
in which case
\[ \imath \dot\psi_l\approx-{1\over 2\omega}\left({\del^2\over\del\rho^2}+{1\over\rho}{\del\over\del\rho}+{2\lambda_l\over (2\rho)^{3\over 2} r_H^{1\over 2}}\right)\psi_l.\]
This absorbs the $\beta^2=r_H^2/r^2$ factor in front of the radial double derivative so as to look more like flat space quantum mechanics, but has an unusual radial power for the angular contribution. Here $\omega$ plays the role of the effective mass and determines the central value
\[ \rho_0={r_H\over 2}\left({m\over \omega}\right)^4\]
around which we wish our approximation to hold.

\subsection{Continuum limit of the discrete black hole}

Here, we send $n\to \infty$ in such a way that the $\Z_n$ geometry becomes $S^1$ with its usual constant metric. The algebraic way to do this was explained in \cite{ArgMa} as a switch from functions on $\Z_n$ to the algebraic circle $\C[s,s^{-1}]$, where classically $s=e^{i\theta}$ for an angle coordinate $\theta$. The limiting calculus is not, however, the classical one on $S^1$, being 2D not 1D. Rather, it is the $q\to 1$ limit of the 2D q-deformed calculus with generators $f^\pm$ and the commutation relations and exterior derivative\cite{ArgMa} 
\[ f^-s=-sf^+,\quad f^+s=s(f^-+(q+q^{-1})f^+),\quad \extd s=sf^+,\quad \extd s^{-1}=s^{-1}f^-.\]
The calculus is inner with
\[ \Theta= {q\over (q-1)^2}(f^++f^-)\]
and has a quantum metric 
\[ g_{S^1}={1\over 2} f^+\tens_s f^- + {q\over (q-1)^2}(f^++f^-)\tens(f^++f^-)\]
 One can check that this is central, i.e. commutes with $s$ and obeys the reality property ${\rm flip}(*\tens*)(g_{S^1})=g_{S^1}$ for a quantum metric if $q$ is real or modulus 1. If we impose $q=e^{2\pi\imath\over n}$ and $s^n=1$ then this is the constant metric ${1\over 2}(q-q^{-1})^2e^+\tens_s e^-$ on $\Z_n$ under the correspondence\cite{ArgMa} 
\begin{equation}\label{ef}  e^\pm={q f^\pm+ f^\mp\over (q-q^{-1})(q-1)}.\end{equation}
In this case  $(q-q^{-1})^2$  is negative, which is the reason for the $-$ sign that was needed in the discrete model. But we do not impose these restrictions and thereby work on the circle.  One still has a flat $*$-preserving QLC with
\[ \nabla f^\pm=0,\quad f^+{}^*=-f^-\]
and $\sigma$ the flip on the basic 1-forms. We now work on $A=C^\infty(\R\times\R_{>0})\tens\C[s,s^{-1}]$ with $t,r,\extd t,\extd r$ classical and graded-commuting with the $s,f^\pm$. We take the metric
\[ g = -{r_H\over r}\extd t \tens \extd t + {r\over r_H} \extd r \tens \extd r +r^2g_{S^1}\]
and we look for QLCs with $\sigma$ assumed to be the flip on the basic 1-forms.

\begin{proposition} The metric $g$ has a canonical Ricci flat $*$-preserving QLC and associated geometry
\begin{align*}	\nabla \extd t =& {1\over 2r} \extd r \tens_s \extd t,\quad \nabla \extd r = -{1\over 2r} \extd r \tens \extd r + {r^2_H\over 2r^3} \extd t \tens \extd t+ r_H g_{S^1},\quad \nabla f^{\pm} =   - \frac{1}{r} \extd r \tens_s f^{\pm},  \\
	{\rm R}_\nabla \extd t =& {1\over r^2} \extd t \wedge \extd r \tens \extd r - {r_H\over 2r} \extd t \wedge g_{S^1},\quad	{\rm R}_\nabla \extd r =-{r_H^2\over r^4}\extd r \wedge \extd t \tens \extd t- {r_H\over 2r}\extd r \wedge g_{S^1},  \\
	{\rm R}_\nabla f^{\pm} =& -{1\over 2 r^2 }f^{\pm} \wedge \extd r \tens \extd r + {r_H^2\over 2r^4}f^{\pm} \wedge \extd t \tens \extd t + {r_H \over r}f^\pm\wedge g_{S^1},\\
	\Delta &=  -{r\over r_H}\del_t^2 + {r_H\over r}\del_r^2 + {r_H\over r^2}\del_r+{1\over r^2}\Delta_{S^1},\quad  \Delta_{S^1}=-{4(1+(q-1)s\del_q)\over (q+1)^2}(s\del_q)^2,\end{align*}
where $\del_q$ is the standard q-derivative so that $\Delta_{S^1}$ on modes $s^l$ has eigenvalue
\[ \lambda_l=- {4q^l[l]_q^2\over (q+1)^2},\quad [l]_q:={1-q^l\over 1-q}.\]
\end{proposition}
\proof First, we can redo the discrete black hole model with $a(r,i)=a r^2$ for any constant factor $a$ for the angular term  $g_{\Z_n}=-ae^+\tens_s e^-$ in the metric. This same factor enters in the connection in the  $\nabla\extd r$ as $g_{\Z_n}$ there. The same happens  for $R_\nabla$  in the term where $e^+\tens_s e^-$ entered. We then replace $g_{\Z_n}$ by $g_{S^1}$ to get the connection as stated, noting that $f^\pm$ are a linear combination of $e^\pm$ so expressions linear in these have the same form.  This version is constructed so as to be isomorphic to the discrete black hole when $q=e^{2\pi\imath\over n}$ and $s^n=1$ are imposed, but these properties do not enter into the computations for a QLC, so this also holds for generic $q$, and likewise for Ricci flatness and for being $*$-preserving when $|q|=1$.  One can also do a direct check of these features and see that $\nabla$ is $*$-preserving also when $q$ is real, as a consequence of $g_{S^1}$ being real in the required sense. 

For Ricci, the antisymmetric lift $i(f^+\wedge f^-)={1\over 2}(f^+\tens f^--f^-\tens f^+)$ of 
\[ f^+\wedge f^-=\left({q-1\over q+1}\right) (q-q^{-1})^2e^+\wedge e^-\]
is equivalent to that of $e^+\wedge e^-$ when we use the correspondence (\ref{ef}). We also use the inverse metric which on the $f^\pm$ comes out as
\[ (f^\pm,f^\pm)=-{4q\over r^2(q+1)^2},\quad (f^\pm,f^\mp)=2{q^2+1\over r^2(q+1)^2}.\]

For the Laplacian, we use $\extd s^l=-{q[l]_q s^l\over q+1}(q[-1-l]_q f^++ [1-l]_q f^-)$ from \cite{ArgMa} and $(\ ,\ )$ to compute $\Delta s^l=(\ ,\ )\nabla\extd s^l=-{4 q^{2+l}\over r^2(q+1)^2}[l]_q^2s^l$, which we write as stated since $s\del_qs^l=[l]_qs^l$ for the standard $q$-derivative $\del_qf(s)=(f(qs)-f(s))/((q-1)s)$. 
The other values of $\Delta$ on functions of $r,t$ are unchanged from the discrete case. In the classical case with $s=e^{\imath l\theta}$, we have $s{\del\over\del s}=-\imath{\del\over\del\theta}$  as the limit of $s\del_q$. \endproof

It remains to say a few words about the actual classical limit of the geometry. As explained in \cite{ArgMa}, this is a joint process $q\to 1$ {\em and} $f^+=-f^-$, with the latter taking precedence so that $g_{S^1}\to -f^+\tens f^+=\extd \theta\tens\extd\theta$ as classically in our normalisation of $g_{S^1}$.  In this way, one arrives as the classical 1+2-dimensional curved metric
\[ g_{class}= -{r_H\over r}\extd t \tens \extd t + {r\over r_H} \extd r \tens \extd r +r^2\extd\theta\tens\extd\theta,\]
which is not, however, Ricci flat. One finds in our conventions (which are $-1/2$ of the usual ones)
\begin{align*} {\rm Ricci}&=-{1\over 2}\left({r_H^2\over 2 r^4}\extd t\tens\extd t- {1\over 2 r^2}\extd r\tens\extd r+ {r_H\over r}\extd\theta\tens\extd\theta\right),\quad S=0,\\
\Delta &=  -{r\over r_H}\del_t^2 + {r_H\over r}\del_r^2 +{1\over r^2}{\del^2\over\del\theta^2}.\end{align*}
The Laplacian agrees with the limit of the $q$-deformed  geometry but Ricci does not. This is due to the 4D cotangent bundle in the quantum model, since the trace gives a different result from the trace in the quotient, where we impose $f^+=-f^-$. Moreover, the dropped terms in the metric that are singular as $q\to 1$ contribute in the calculation of ${\rm Ricci}=0$ in the quantum model. 

\subsection{Discrete black hole model for $\beta(r)<0$}

Here we briefly analyse the case where $r_H<0$ in our previous presentation of the discrete black hole. More precisely, we still define $r_H=2GM>0$ but replace $r_H$ by $-r_H$ and we also replace $t$ by $r$ and $r$ by $t$ in all the formulae (\ref{eq:metric_pol_rii_flat})-(\ref{disclap}) so as the match the signature. Thus, the quantum metric and resulting quantum geometry are now
\begin{align*}g &= -\frac{t}{r_H} \extd t \tens \extd t + \frac{r_H}{t} \extd r \tens \extd r - t^2 e^+\tens_s e^-,\\
	\nabla \extd r &= {1\over 2t} \extd r \tens_s \extd t,\quad \nabla \extd t = -{1\over 2t} \extd t \tens \extd t + {r^2_H\over 2t^3} \extd r \tens \extd r- r_H e^+\tens_s e^-, \\
	\nabla e^{\pm} &=   - \frac{1}{t} \extd r \tens_s e^{\pm},  \quad \Delta = -{r_H\over t}\del_t^2 + {t\over r_H}\del_r^2 - {r_H\over t^2}\del_t + {2\over t^2}(\del_+ + \del_-)
\end{align*}
with a curvature singularity now at $t=0$. We next make a change of  variable 
\[ t = ({3\tau\over 2})^{\frac 2 3}r_H^{\frac 1 3}= \eta(\tau)^2 r_H,\quad \eta(\tau)=\left({3\tau\over 2 r_H}\right)^{\frac 1 3}\]
 in order to have a constant term in the `time` coefficient of the metric, so that the quatum geometric structures become
\begin{align*}g &= - \extd \tau \tens \extd\tau + \eta^{-2} \extd r \tens \extd r - \eta^4 r_H^2 e^+\tens_se^-,\\
	\nabla e^\pm &= -{2\over 3\tau} \extd \tau \tens_s e^\pm, \quad  \nabla\extd \tau = -{1\over 3\eta^2\tau} \extd r \tens \extd r - \eta r_H e^+\tens_se^-,\\
	\nabla \extd r &= {1\over 3\tau}\extd r \tens_s \extd \tau, \quad \Delta = - \del_\tau^2 + {1\over 3\tau}\del_\tau + \eta^2\del_r^2  +{ 2\over \eta^4 r_H^2}(\del_+ + \del_-). 
\end{align*}

We now do the parallel analysis to Section~\ref{kg_BhPoly}. Using the above Laplacian for the Klein-Gordon equation, we first look for solutions of the form $\phi=e^{-\imath m \tau}\psi_l(\tau,r)$ where $\psi_l$ is slowly varying in $\tau$ and with eigenvalue $\lambda_l$ for the angular sector. Ignoring $\ddot\psi_l$, we have
\[ \imath\dot\psi_l=-{\eta^2\over 2m-{\imath\over 3 \tau}}\left(\del_r^2+{8 \lambda_l\over 9\tau^2}\right)\psi_l,\]
where dot denotes $\del_\tau$. If we assume that we are
very far from the $\tau=0$ singularity in the sense
\[ \tau >> {1\over m}\]
(i.e. at macroscopic times much larger than the Compton wavelength in time units), we have
\begin{equation}\label{expsch}\imath \dot\psi_l\approx - {\eta^2\over 2m} \left(\del_r^2+  {8\lambda_l \over 9\tau^2} \right)\psi_l.\end{equation}
This looks, as expected, a bit like quantum mechanics, not in the presence of a point source potential but rather with an overall time-dependent expansion factor and a time-dependent contribution of the angular momentum.  Note that $e^{-\imath m \tau}$ does not itself obey the Klein Gordon equation.

Next, we look for the `comoving' behaviour, noting that solutions of the Klein-Gordon equation of mass $m$ and $l=0$ are in fact given by  Hankel functions, of which we focus on the first type, 
\[ \phi_m(\tau)=\tau^{2\over 3}H_{2\over 3}^{(1)}(m\tau). \]
Here, the real and imaginary parts (Bessel J, K functions respectively) oscillate, $\phi_m(0)$ is a nonzero (imaginary) value and $|\phi_m|^2$ gradually increases with time.  This therefore plays the role of an exact plane wave. Relative to this, we look for solutions of the form 
\[ \phi(\tau,r)=\phi_m(\tau)\psi_l(\tau,r)\]
with $\psi_l$ slowly varying in $\tau$, leading to a Schroedinger-like equation 
\[  \imath\dot\psi_l= -{\eta^2  h(m\tau)\over 2m} \left(   \del_r^2 +  {8 \lambda_l\over 9\tau^2}  \right)\psi_l,\]
where 
\[h(s)=\imath {H_{\frac{2}{3}}^{(1)}(s)\over   H_{-\frac{1}{3}}^{(1)}(s)-{1\over 6 s} H_{\frac{2}{3}}^{(1)}(s)}\approx 1\]
 \begin{figure}
	\includegraphics[scale=.75]{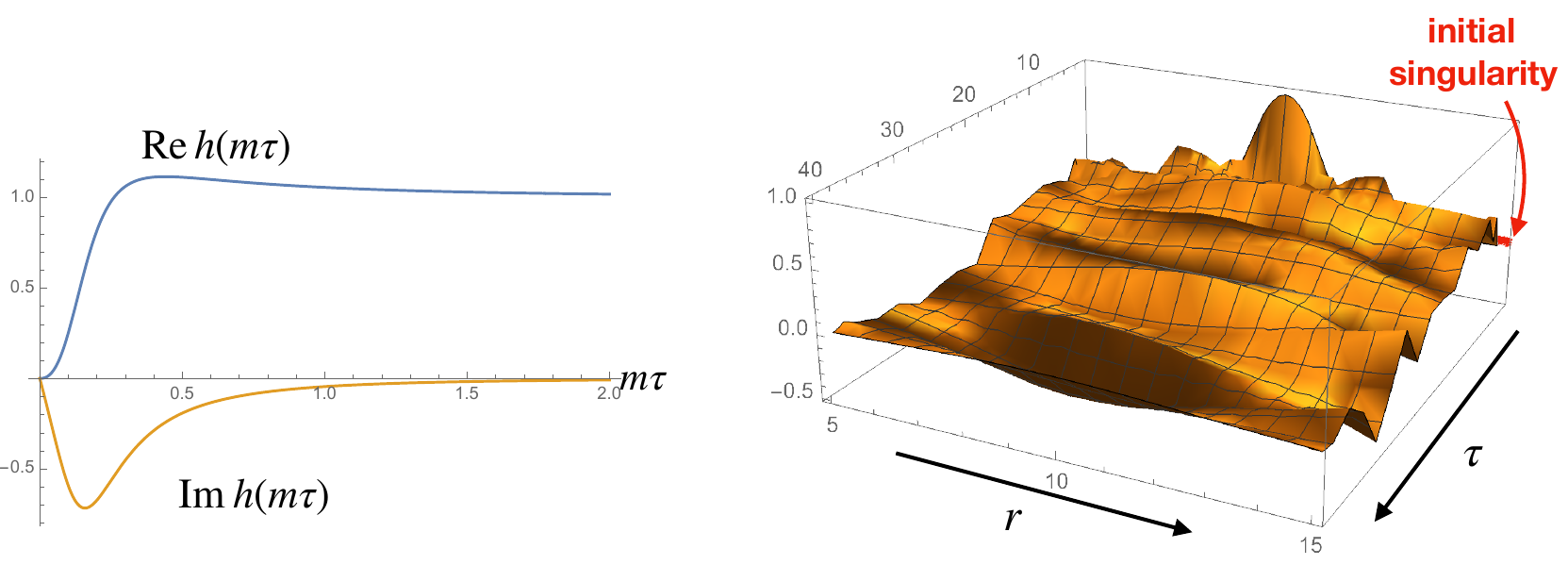} 
	\caption{Function $h(m\tau)$ in definition of Schroedinger-like equation for discrete black hole metric and evolution of a Gaussian centred at $r=10r_H$ at $\tau=1/m$, for $r_H=m=1$ and $l=0$. The essentially zero initial values at $r=0,20r_H$ are held fixed. \label{figha}}\end{figure}
\noindent for large $s$, as shown on the left in Figure~{\ref{figha}. Here, one can see that $h(m\tau)$  approaches 1 very rapidly as $\tau>>1/m$. In other words, the behaviour near the $\tau=0$ singularity is different but for larger $\tau$ the effective Schroedinger-like equation is now much more sharply approximated by 
(\ref{expsch}) than before.  
 
The numerical solution for the real part of these equation is shown on the right in Figure \ref{figha}, where we used the exact function $h(m\tau)$ and set the initial Gaussian at $m\tau=1$. The evolution becomes noticeably constant in $r$ compared to regular quantum mechanics.  Some of the noise in the picture comes from the numerical approximation.

\section{Concluding remarks}

We have solved for the quantum Levi-Civita connection and hence found the quantum geometry for 
quantum metrics with each sphere at radius $r,t$ replaced by a fuzzy sphere $\C_\lambda[S^2]$. We did this for both FLRW-type metrics (\ref{eq:general_metric}) and static black-hole like metrics (\ref{genSchlikeg}) in polar coordinates and general metric $g_{ij}$ on the fuzzy sphere. We also completed the discrete case with static metric (\ref{eq:bhm-pol}), where each sphere is replaced by $\Z_n$ as a discrete circle or its noncommutative $S^1$ limit, the FLRW-like case having already been treated in \cite{ArgMa}. After the general analysis, we specialised to the constant coefficient or `round' metric $g_{ij}=k\delta_{ij}$ in the fuzzy case and regular polygon metric $-e^+\tens_s e^-$ in the discrete case, respectively, and solved the Friedmann equations for the cosmological model and the Ricci=0 equation for the black-hole-like models. 

The four models between them show a remarkably consistent `dimension jump' phenomenon where the radial-time sector behaves as for a classical model of one dimension higher. The origin of this from a mathematical point of view is what has been called a `quantum anomaly for differential structures' \cite{Ma:spo,BegMa:qua,FreMa}, where quantisation of an algebra while preserving symmetries typically has an obstruction requiring either a breakdown of associativity or, which is our approach here, an extra cotangent dimension. This then affects both the Ricci tensor and Laplace operator, which is not surprising, but it is remarkable the result appears so simply as a classical dimension jump. The consequence from a physical point of view is striking: if each sphere at $r,t$ is better modelled as fuzzy due to quantum gravity effects, which is plausible enough if one wanted to preserve rotational symmetry but allow for some noncommutativity of spatial coordinates, then Ricci flat solutions, in particular, have a very different long range behaviour in 4D, being now of the form of a 5D black hole with the black hole appearing as a source of an inverse cubic gravitational force. In the discrete circle case, as well as in its noncommutative circle limit, the fact that the circle has zero constant curvature in contrast to $S^2$ also resulted in dropping the $1$ in the usual Schwarzschild factor $\beta=1-2GM/r$, which meant that we only approximated the inside of a black hole far from the horizon. We are not proposing the model as 4D physics since the angular sector remains a circle not a sphere but it could be of interest in 2+1 gravity and meanwhile it illustrates that it is possible to have a nonflat Ricci=0 quantum geometry in 3D, ultimately because of the hidden extra cotangent direction. The geometric meaning of the extra dimension was discussed in \cite{ArgMa} as a kind of normal to the circle but without actually extending the circle to an ambient plane. 

In summary, we offer new models with different radial-time behaviour from those expected. We do not know if such effects could be relevant to real world cosmology but the idea of modified gravity\cite{Mil} is not new and it is possible that this new theoretical phenomenon could be of interest. We also introduced a novel  `comoving' Schroedinger-like equation i.e. slowly varying relative to an actual solution $\phi_m$ of the Klein-Gordon equation. We have not developed this as a formal theory but this could certainly be looked at further as a complement to  more established methods of quantum field theory on curved spaces\cite{Birrel,ParTom, MukWin}. In particular, the solutions $\psi_l$ appear in practice to dissipate over time even for a regular black hole background. This could potentially relate to ideas for gravitational measurement, but note that this would be a very different phenomenon from gravitational decoherence\cite{Bas}, which applies to density matrices not pure states. 

Of course, our analysis is only as good as the assumed formalism, and here we assumed the constructive approach to quantum Riemannian geometry as in \cite{BegMa}. As in the concluding remarks in \cite{ArgMa}, it would be fair to say that the Einstein tensor in the general set up is not known and the proposal for Ricci is merely by analogy (a trace of Riemann) rather than springing from a more conceptual understanding. In general, in order to take a trace, the formulation of Ricci depends on a lifting map $i:\Omega^2\to \Omega^1\tens_A\Omega^1$ which classically would express a 2-form as an antisymmetric tensor but which in general depends on the structure of $\Omega^2$. Fortunately, for the models in the present paper, as in \cite{ArgMa}, there are natural  basic 1-forms with respect to which $\Omega^2$ is given by skew-symmetrising, so we can take $i$ in the standard form as classically. We also found for the FLRW model (\ref{flrwEins}) and for the spatial geometry of the fuzzy black hole model (Proposition~\ref{fuzBHspatial}), that the quantum Einstein tensor defined by ${\rm Eins}={\rm Ricci}-{S\over 2}g$, where $S$ is the Ricci scalar, led as expected to $\nabla\cdot{\rm Eins}=0$. 

The physics of such a quantum Ricci and Einstein tensor remains, however,  to be understood much better. For example, we took the view for the fuzzy black hole that an observer sees the event horizon at  $r=r_H$, which is the physical parameter, but equated it to $2GM$ for an effective `Schwarzschild mass' for the purposes of comparison. To do better, one should have a  noncommutative version of ADM theory, but we saw that naively adopting its classical physical formulation in terms of the spatial Einstein tensor\cite{Ash,Cru,Mia}  but using the spatial quantum Einstein tensor gave an infinite ADM mass as a consequence of the dimension jump. It is also the case that the models in the present paper do not concern quantum gravity itself but rather noncommutative classical gravity proposed to model better quantum gravity effects. It remains to understand mechanisms for how our class of models could indeed emerge from an underlying theory. Thus, \cite{Hoo,FriMa} gave some reasons for why the  fuzzy sphere could emerge from 2+1 quantum gravity, but it is unclear how such arguments might extend to the higher dimensional models proposed. By contrast, \cite{Ale} studies effects on the interior of a black hole from loop quantum gravity, but the considerations there are quite different.

Nevertheless, the class of models studied in this paper were particularly nice as far as the quantum geometry itself is concerned and more tractable than fully noncommutative models where $r,t$ need not be classical as they were for us. We refer to the concluding remarks of \cite{ArgMa} for a wider discussion. Also remaining, even for our simple class of models, is to study quantum geodesics using the Schroedinger-like formalism of \cite{Beg:geo, BegMa:geo}. This requires further machinery, notably the construction of a certain $A$-$B$-bimodule connection (where $B$ is the classical geodesic time algebra), and will be considered elsewhere. These are some direction as we see it for further work.

\end{document}